\documentclass[preprint,aps,pra,showpacs]{revtex4}
\usepackage{tabularx}
\usepackage{dcolumn}
\usepackage{graphics}
\usepackage{bm}
\usepackage[dvips]{graphicx,xcolor}
\usepackage[dvips]{rotating}
\usepackage[latin1]{inputenc}
\usepackage{dcolumn}
\usepackage{tabularx}
\usepackage{amsmath}
\usepackage{amssymb}
\usepackage{textcomp}
\usepackage[section]{placeins}

\begin{document}
\draft

\title{Photodetachment near an attractive force center}

\author{X. P. You and  M. L. Du }\email{
duml@itp.ac.cn}
 \affiliation{Institute
of Theoretical Physics, Chinese Academy of Sciences,
Beijing 100080, China}



\date{\today}

\begin{abstract}
This article studies the photodetachment of a single electron anion near an attractive center.
Both the differential and total photodetachment cross section are analysed.
We obtain the electron flux crossing through a spherical  detector centered at the force center using the semiclassical approximation.
The closed-orbit theory gives the total cross section which contains a smooth background and an oscillatory part.
Concrete calculations and discussions are carried out for two types of wave source: the $s$- and $p_z$-wave source.
Photodetachment processes for three conditions are compared: an anion near an attractive center, near a repulsive center and in a homogeneous electric field.

\par

\pacs{32.80Gc,32.60.+i,31.15.xg}

\end{abstract}

 \maketitle


\section{Introduction}

Given some special external fields,
photodetachment of anions and photoionization of atoms has been studied using different theoretical and experimental methods.
Demkov $et$ $al$ \cite{Demkov} and Fabrikant \cite{Fabrikant1}
have analysed the processes of photoionization and photodetachment in homogeneous electric fields.
The oscillatory structure of  the cross section was predicted.
Using $\mathrm{\pi}$-  polarized laser light,
Bryant $et$ $al$ observed ripplelike structures in the photodetachment cross section of $\rm{H^-}$ near threshold in motional electric fields
\cite{Bryant, Bryant2}.
The explanations of oscillations and corresponding calculation methods have been presented by Rau $et$ $al$\cite{Rau, Rau2} and by Du $et$ $al$\cite{Du1, Du2, Du3}.

In the past years, photodetachment has been studied for different anions \cite{Gibson,Gibson2} in different imposed external fields \cite{Du4, Peters1, Peters2}.
In $2008$, Xing $et$ $al$ observed photoelectron angular distributions for a series of dicarboxylate dianions,
 $^-\rm{O_2C(CH_2)_nCO_2}^-$, using photoelectron imaging \cite{Xiaopeng Xing}.
 The distributions are shown to be dominated by the interplay between intramolecular Coulomb repulsion and
 the detached electron kinetic energies.
 Then Yang, Delos and Du explained the experimental results by constructing a theoretical  model of photodetachment near a repulsive center.
 They have considered the effects of the nearby repulsive center on both the differential \cite{B. C. Yang} and total
 \cite{B. C. Yang2} photodetachment cross section.
It has been observed in experiments that a nearby attractive central force can also influence the cross section.
In 1991,
Swenson and Burgd$\rm\ddot{o}$fer $et$ $al$ observed oscillation structures in autoionizing spectra for low-energy $\rm{He^++He}$ collisions\cite{Swenson}.
Here, we study the photodetachment of a single electron anion near an attractive center in this article.

  This article is organized as follows.
  In Sec. {\uppercase\expandafter{\romannumeral2}, the theoretical model is described and the detached electron
  motion analysed.
  In Sec. {\uppercase\expandafter{\romannumeral3},
  we use the semiclassical approximation to derive the detached electron flux crossing a spherical detection surface.
  Several representative cases are discussed for $s$-wave sources and $p_z$-wave sources.
  Using the closed-orbit theory \cite{Du and Delos},
  we obtain the total phototodetament cross section in Sec. {\uppercase\expandafter{\romannumeral4},
  where some calculation results are displayed through the field-induced modulation function and reduced total cross section.
  In addition, we study the photodetachment of $\rm{H}^-$ near an attractive center.
  Comparisons are carried out for $\rm{H^-}$ near an attractive center, $\rm{H^-}$ near a
  repulsive center and $\rm{H^-}$ in a uniform electric field.
  The results are then generalized to the photodetachment of common anions.
  Finally, one can find the brief conclusions in Sec. {\uppercase\expandafter{\romannumeral4}.
  Atomic units will be used unless otherwise explicitly indicated.

\section{theoretical model}

We consider the following theoretical model: the photodetachment of a single electron anion near a cation (Fig.1).
The anion is assumed to be monatomic.
It is the source of electron waves.
The cation with the charge number $\alpha$ provides the detached electron with a central attractive force.
The $z$ axis is set from the attractive force center (the cation) to the source point (the anion).
The photodetament process has been discussed in previous literatures \cite{Du1,Peters1}.
Here we present it briefly.
Initiously the electron is loosely bounded to the atom by a short range potential.
When laser light is applied, the ion may absorb a photon and detach an electron.
Then the electron escapes from the source region and moves under the influence of the attractive force.

For convenience, the electron motion is described using spherical coordinates $(r,\theta,\varphi)$ relative to the force center (Fig. 1).
Due to the cylindrical symmetry of the system, $\varphi$ component can be omitted.
The Hamiltonian for the detached electron is
\begin{equation}\label{1}
    H=\frac{p_r^2}{2}+\frac{p_\theta^2}{2r^2}-\frac{\alpha}{r}+\frac{\alpha}{d},
\end{equation}
where d is the distance between the source point and the cation.
To make the potential energy equal
zero at the source, the constant $\frac{\alpha}{d}$ is added.

The corresponding orbit equation is needed for the analysis of the electron motion.
Similar to the system with a repulsive force center \cite{B. C. Yang}, the scaled energy $\overset{\sim}E=Ed/{\alpha}$ is used to make the analysis easier.
$E$ denotes the electron energy.
Two complementary parameters $\xi$ and $\eta$ are also used:
\begin{equation}\label{2}
  \xi^2=\frac{2p_0^2}{p_0^2-p_\infty^2}=2\overset{\sim}E,
\end{equation}
\begin{equation}\label{3}
  \eta^2=\frac{2p_\infty^2}{p_0^2-p_\infty^2}=2(\overset{\sim}E-1).
\end{equation}
$p_0$ is the initial momentum of the electron after detachment and $p_\infty$ the momentum when the electron
 goes infinitely far. If $E$ is less than $\alpha/d$,
 the electron motion is bounded by the attractive force provided by the cation.
 This is out of our consideration. So the scaled energy $\overset{\sim}E$ is larger than $1$.
  However for a system with a repulsive center, there is no such limitation \cite{B. C. Yang}.
 The electron trajectory is hyperbolic. Fig. 2 gives some trajectories starting with different emergence angles.
 Using classical mechanics, the orbit equation of the electron can be derived:
 \begin{eqnarray}
   r=\left(\frac{\xi^2\sin^2\beta}{\varepsilon\cos(\theta-\theta_0)+1}\right)d, & \overset{\sim}E>1,
 \end{eqnarray}
where $\beta$ is the emergence angle and $\theta_0$ the perihelion angle.
$\varepsilon$ is the eccentricity and equals $\sqrt{1+\xi^2\eta^2\sin^2\beta}$.
\begin{equation}
  \theta_0=\left\{\begin{array}{lll}
  -\arccos\left(\frac{\xi^2\sin^2\beta-1}{\varepsilon}\right) & if &\beta\le\frac{\mathrm{\pi}}{2},\\
  \arccos\left(\frac{\xi^2\sin^2\beta-1}{\varepsilon}\right) & if &\beta>\frac{\mathrm{\pi}}{2}.
  \end{array}
  \right.
\end{equation}
Assume a spherical detector centered at the origin (the force center).
Given the detector radius and emergence angle,
one can calculate the detection angle $\theta$ according to the orbit equation.
$\theta$ has the form
\begin{eqnarray}
  \theta=\theta_0+\arccos\left(\frac{\frac{d}{r}\xi^2\sin^2\beta-1}{\varepsilon}\right),&r>d.
\end{eqnarray}
 When r is infinitely large, $\theta$ has an asymptotic value $\theta_\infty$ which reads
 \begin{equation}
   \theta_\infty=\theta_0+\arccos(-\frac{1}{\varepsilon}).
 \end{equation}
 There is another special detection angle denoted as $\theta_c$.
 $\theta_c$ is calculated when the emergence angle $\beta$ equals $\pi/2$.
 It can be expressed as
 \begin{equation}
   \theta_c=\arccos\left(\frac{2\overset{\sim}E\frac{d}{r}-1}{2\overset{\sim}E-1}\right).
 \end{equation}
 For a point $(r,\theta)$ on the detector,
 there are always two trajectories from the source point to the detection point.
 They are distinguished using their emergence angle $\beta_1$ and $\beta_2$.
 The index $1$ denotes the trajectories which always eject upward.
 The expressions for $\beta_1$ and $\beta_2$ are obtained from the orbit equation Eq. $(4)$.
 \begin{equation}
   \beta_1=\mathrm{\pi}-\arcsin{(\sqrt{x_1})}
 \end{equation}
 with
 \begin{equation}
   x_1=(1-\cos\theta)\frac{\frac{\overset{\sim}E-1}{2}\left(\sqrt{1+\cos\theta}-\sqrt{\cos\theta+\frac{\overset{\sim}E^2+2\overset{\sim}E\frac{d}{r}-1}{(\overset{\sim}E-1)^2}}\right)^2-\frac{1}{\overset{\sim}E-1}\frac{d}{r}}{(2\overset{\sim}E-1)(1-\frac{d}{r})(1-\cos\theta_c)+4\overset{\sim}E\frac{d}{r}(1-\cos\theta)}.
 \end{equation}
\begin{equation}
  \beta_2=\left\{
  \begin{array}{lll}
  \arcsin(\sqrt{x_2})&if&0\le{\theta}\le{\theta}_c\\
  \mathrm\pi-\arcsin(\sqrt{x_2})&if&{{\theta}_c}<{\theta}\le\mathrm{{\pi}}
  \end{array}
  \right.
\end{equation}
with
\begin{equation}
   x_2=(1-\cos\theta)\frac{\frac{\overset{\sim}E-1}{2}\left(\sqrt{1+\cos\theta}+\sqrt{\cos\theta+\frac{\overset{\sim}E^2+2\overset{\sim}E\frac{d}{r}-1}{(\overset{\sim}E-1)^2}}\right)^2-\frac{1}{\overset{\sim}E-1}\frac{d}{r}}{(2\overset{\sim}E-1)(1-\frac{d}{r})(1-\cos\theta_c)+4\overset{\sim}E\frac{d}{r}(1-\cos\theta)}.
\end{equation}

Each point $(r,\theta)$ on the detector can be reached through two paths from the source point.
That is to say each detection angle $\theta$ corresponds to two emergence angles: $\beta_1$ and $\beta_2$.
Fig. 3 shows this kind of relationship.
Note that the two trajectories come from different sides of the $z$ axis.
In Fig. 2, the red lines only intersect with the blue lines, and vice versa.
Unlike the system with a repulsive force center \cite{B. C. Yang},
any point on the detector can be reached for that with the attractive force center and there is no forbidden region.
However for trajectories with an emergence angle $\beta_1$,
the Maslov index equals $1$, not $0$.
This will be discussed later in Sec. A of Sec. {\uppercase\expandafter{\romannumeral3}.
\section{differential cross section}

The differential cross section has the following formula
\begin{equation}
  \frac{\mathrm{d}\sigma(\bm{q})}{\mathrm{d}s}=\frac{2\pi{E_{ph}}}{c}\bm{j\cdot{n}}.
\end{equation}
 $E_{ph}$ represents the photon energy, $c$ the light speed,
$\bm{j}$ the electron flux. $\mathrm{d}s$ is the area element of the surface which $\bm{j}$ crosses
and $\bm{n}$ is the unit normal vector of the surface.
$\bm{q}$ is the coordinates of the detection point.
 For the spherical detection surface centered at the origin,
the above formula becomes \cite{B. C. Yang}:
\begin{equation}
  \frac{\mathrm{d}^2\sigma(r,\theta,\varphi)}{r^2\sin{\theta}\mathrm{d}{\theta}\mathrm{d}{\varphi}}=\frac{2\pi{E_{ph}}}{c}j_r
\end{equation}
where $j_r$ is the component of $\bm{j}$ along the radius of the spherical detector.
\begin{equation}
  j_r=\mathrm{Im}(\psi^*\frac{\partial\psi}{\partial{r}})
\end{equation}
$\psi$ is the wave function of the detached electron reaching the detection point $(r,\theta,\varphi)$.
\subsection{Electron wave function}

The electron wave function $\psi$ can be constructed semiclassically \cite{Du and Delos,Gutzwiller, Delos}.
After photodetachment,
the electron goes out leaving the residual atom behind.
Initially, the outgoing wave function ${\psi}_{\rm{out}}$ satisfies the inhomogeneous Shr\"odinger equation \cite{Du2}
\begin{equation}
  (E+\frac{1}{2}\nabla^2_s-V){\psi}_{\rm{out}}=D{\psi_{\mathrm{i}}}.
\end{equation}
The subscript $s$ denotes the coordinates relative to the source point.
$V$ is the short range potential provided by the neutral atom.
$D$ is the dipole operator.
${\psi}_{\rm{i}}$ represents the initial bound state.
Away from the source region, the short range potential can be ignored
and far away enough the electron only moves under the influence of the attractive force.
Assume a selected spherical surface $\Gamma$ centered at the residual atom.
Its radius R is in an appropriate range
where the effect of the attractive force is so small that can be neglected and the use of the asymptotic wave function can give a perfect result.
Hence it satisfies the inequality
\begin{equation}
  {\frac{1}{k}}\ll{R}\ll{\frac{\overset{\sim}E}{\overset{\sim}E-1}d}.
\end{equation}
$k$ is the magnitude of the initial wave vector.
 $k=\sqrt{2E}$.
 On the spherical surface $\Gamma$, the outgoing wave function may be written as \cite{B. C. Yang}:
\begin{equation}
  \psi_{\rm{out}}(R,\beta,\phi)=C(k)Y_{lm}(\beta,\phi)\frac{e^{\mathrm{i}kR}}{R},
\end{equation}
where $(R,\beta,\phi)$ are spherical coordinates relative to the source point,
$C(k)$ is a factor dependent on specific conditions, $Y_{lm}(\beta,\phi)$ the spherical harmonic function.

When the electron goes out further, the effect of the attractive force comes into consideration.
The orbit of the electron has been discussed in Sec. $\mathrm{\uppercase\expandafter{\romannumeral2}}$.
The electron wave propagates from the surface $\Gamma$ to the detection point $(r,\theta,\varphi)$ semiclassically. The wave function at $(r,\theta,\varphi)$ can be obtained using semiclassical approximation.
The formula is
\begin{equation}
  \psi(r,\theta,\varphi)=\sum_{\nu}\psi_\mathrm{out}(R,\beta_{\nu},\varphi)A_{\nu}\exp[\mathrm{i}(S_{\nu}-\mu_{\nu}\frac{\pi}{2})].
\end{equation}
The index $\nu$ is used to designate different trajectories.
$A_{\nu}$ is the probability amplitude which indicates the divergence of adjacent trajectories.
$S_{\nu}$ is the action along the corresponding electron trajectory.
$\mu_{\nu}$ is the Maslov index for the path going from the initial point $(R,\beta_{\nu},\varphi)$ to the final point $(r,\theta,\varphi)$.
Formulas for the physical quantities above will be given thereinafter.

$A_{\nu}$ equals
\begin{equation}
  A_{\nu}=\left|\frac{J(t=0)}{J(t=T)}\right|^{1/2},
\end{equation}
where $J(t)$ is the Jacobian determinant at time $t$ \cite{Delos}.
 When $t=0$,
 \begin{equation}
   J(t=0)=kR^2\sin\beta_\nu.
 \end{equation}
For this system which has a force center, the Jacobian determinant at time $t$ can be reduced to one dimension\cite{B. C. Yang}:
\begin{equation}
  J(t)=p_{\nu{r}}r^2\sin\theta\left(\frac{\partial\theta}{\partial\beta_\nu}\right)_r.
\end{equation}
 $p_{\nu{r}}$ represents the radial momentum of the electron with the emergence angle $\beta_{\nu}$.
A detailed reduction process can be found in Ref. $[16]$.
 Combine Eq. $(20)$, Eq. $(21)$ and Eq. $(22)$, one gets
 \begin{equation}
   A_\nu=\left|\frac{kR^2\sin\beta_\nu}{p_{\nu{r}}r^2\sin\theta\left(\frac{\partial\theta}{\partial\beta_\nu}\right)_r}\right|^{1/2}.
 \end{equation}
 After careful calculation, we get $\left(\frac{\partial\theta}{\partial\beta_\nu}\right)_r$ which will be used subsequently:
 \begin{equation}
   \left(\frac{\partial\theta}{\partial\beta_\nu}\right)_r=1+\frac{1+\eta^2}{\varepsilon^2}-\frac{\eta^2+\frac{d}{r}(1+\varepsilon^2)}{\varepsilon^2}\times\frac{k}{p_{\nu{r}}}\cos\beta_\nu.
 \end{equation}

 $S_\nu$ here refers to Hamilton's characteristic function and would be obatined from the expression
 \begin{equation}
   S_\nu=\int_\nu\bm{p\cdot}\mathrm{d}\bm{q}.
 \end{equation}
 Given a detector radius $r$ and an emergence angle $\beta$,
 the trajectory is determined and the detection angle $\theta$ is known.
 Given $r$, $\theta$ and which kind of the trajectories the electron travels along (trajectory ``$1$'' or ``$2$''),
 the orbit is also determined. It follows that \cite{B. C. Yang}
 \begin{equation}
   \left(\frac{\partial{S(r,\beta)}}{\partial\beta}\right)_r=\left(\frac{\partial{S(r,\theta)}}{\partial\theta}\right)_r\left(\frac{\partial{\theta(r,\beta)}}{\partial\beta}\right)_r=L\left(\frac{\partial{\theta(r,\beta)}}{\partial\beta}\right)_r.
 \end{equation}
 $L=kd\sin\beta$ is the conserved angular momentum relative to the force center.
 Eq. $(25)$ can be integrated to give
 \begin{equation}
   S(r,\beta_\nu)=S(r,0)+S_d
 \end{equation}
 with
 \begin{eqnarray}
    {\nonumber}S_d&=&\int_0^{\beta_\nu}{kd}\sin\beta\left(\frac{\partial\theta}{\partial\beta}\right)_r{\rm{d}}\beta \\
       &=&kd\left(1-\cos\beta_\nu+\frac{1}{\xi\eta}\ln\frac{1+\gamma}{1+\gamma\cos\beta_\nu}+\frac{1}{\xi\eta}\ln\frac{\eta^2+\frac{d}{r}+\frac{\xi\eta}{k}p_{{\nu}r}}{\eta^2+\frac{d}{r}+\frac{\xi\eta}{k}p_{0r}}-\frac{\frac{kd}{r}\sin^2\beta_\nu}{p_{0r}+p_{{\nu}r}}\right)
     \end{eqnarray}
 in which
 \begin{equation}
   \gamma=\frac{2\xi\eta}{\xi^2+\eta^2}.
 \end{equation}
 For this system, there are only two trajectories as discussed before.
 So the wave function $\psi$ can be written in the form
 \begin{equation}
   \psi(r,\theta,\varphi)=\psi_{\rm{out}}(R,\beta_1,\varphi)A_1\exp\left[\mathrm{i}\left(S_1-\mu_1\frac{\mathrm\pi}{2}\right)\right]+\psi_{\rm{out}}(R,\beta_2,\varphi)A_2\exp\left[\mathrm{i}\left(S_2-\mu_2\frac{\mathrm{\pi}}{2}\right)\right].
 \end{equation}

 Each time a trajectory passes through a caustic,
 the corresponding Maslov index increases by 1 and the wave function undergoes a phase loss of $\mathrm\pi/2$.
 Although there is no forbidden region in the system (Fig. 2),
 a caustic still exists.
 From the discussion given in Sec. $\mathrm{\uppercase\expandafter{\romannumeral2}}$,
 one can see that trajectories starting with
 emergence angle $\beta_1$ (trajectory ``1'') always cross the symmetry axis which connects the force center and the source point.
 Different trajectories with the same $\beta_1$ but different $\varphi$ all converge onto the axis.
 So the region near the axis is a singular region.
 When a trajectory crosses it, the corresponding Maslov index increases by $1$.
 The trajectory ``1'' crosses the axis once, so $\mu_1=1$, while the trajectory ``2''
 never crosses it and $\mu_2=0$.

\subsection{Electron flux}

The behavior of the electron flux $j_r$ reflects the photoelectron angular distribution
which can be observed in experiments \cite{Xiaopeng Xing, Blondel1}.
Substitute Eq. $(30)$ to Eq. $(15)$, the electron flux ${j_r}$ becomes
\begin{eqnarray}
  {\nonumber}j_r&=&\frac{k}{r^2}\times{|C(k)|}^2\times\Big\{\mathcal{A}_1^2{|Y_{lm}(\beta_1,\varphi)|}^2+\mathcal{A}_2^2{|Y_{lm}(\beta_2,\varphi)|}^2\\
  {\nonumber}&&+\frac{p_{1r}+p_{2r}}{\sqrt{p_{1r}p_{2r}}}\mathcal{A}_1\mathcal{A}_2Y_{lm}(\beta_1,\varphi)Y_{lm}(\beta_2,\varphi)\\
  &&\times\cos\left(\Delta{S}-\frac{\pi}{2}\right)\Big\}
\end{eqnarray}
with
\begin{equation}
  \mathcal{A}_\nu=\left|\frac{\sin\beta_\nu}{\sin\theta\left(\frac{\partial\theta}{\partial\beta_\nu}\right)_r}\right|^{1/2}
\end{equation}
and
\begin{equation}
  \Delta{S}=S_1-S_2.
\end{equation}

During calculation, the reduced flux defined in Ref. $[16]$ is used:
\begin{equation}
  \overset{\sim}{j_r}=j_rr^2\times\frac{1}{k}\times\frac{1}{|C(k)|^2}\times\frac{1}{N_{lm}^2},
\end{equation}
where $N_{lm}$ is the normalization coefficient of the spherical harmonic function.
In this article, we consider two types of wave source: the $s$-wave source and $p_z$-wave source.
In experiments, an $s$-wave can be produced  by $\rm{Br}^-$, $\rm{^{16}O^-}$, $\rm{Cl}^-$ etc \cite{Blondel1, Gibson}
and a $p$-wave by $\rm{H}^-$, $\rm{Au}^-$ etc \cite{Blondel2, Gibson}.
As for the $p_z$-wave source mentioned above,
the subscript $z$ implies the application of linearly polarized laser light along the z axis (the symmetry axis of the system).

The reduced flux $\overset{\sim}{j_r}$ is
 \begin{equation}
   \overset{\sim}{j_r}=\mathcal{A}_1^2+\mathcal{A}_2^2+\frac{p_{1r}+p_{2r}}{\sqrt{p_{1r}p_{2r}}}\mathcal{A}_1\mathcal{A}_2\cos\left(\Delta{S}-\frac{\pi}{2}\right)
 \end{equation}
for $s$-wave photodetachment and
\begin{equation}
  \overset{\sim}{j_r}=\mathcal{A}_1^2\cos^2\beta_1+\mathcal{A}_2^2\cos^2\beta_2+\frac{p_{1r}+p_{2r}}{\sqrt{p_{1r}p_{2r}}}\mathcal{A}_1\mathcal{A}_2\cos\beta_1\cos\beta_2\cos\left(\Delta{S}-\frac{\pi}{2}\right)
\end{equation}
for $p_z$-wave photodetachment.

\subsection{Calculations and discussions}

Because of the existence of the two trajectories from the source point to the detection point,
interference happens between the two electron wave functions.
So oscillatory structures can be found in the photoelectron spatial distribution pattern.
Note that for the photodetachment near a repulsive center,
the range of the detection angular $\theta$ runs from $0$ to a definite angle before $\pi$.
The quantum tunneling effect makes this angle larger than ${\theta}_m$ (the maximum value of theta calculated from the orbit equation)\cite{B. C. Yang}.
Whereas for the current system we study,
the range of $\theta$ runs from $0$ to $\pi$.
That is, the emitted electron runs through the whole spherical surface.
\subsubsection{The effect of the detector radius}

First we study the effect of the detector radius.
Fig. 4 depicts the variation of the reduced electron flux $\overset{\sim}{j_r}$ with increasing detector radii for $s$-wave photodetachment.
The increase of $r$ has a strong effect on the differential cross section both near $\theta=0$ and $\theta=\pi$.
when one increases $r$, the oscillation amplitude around $\theta=0$ decreases and that around $\theta=\pi$ increases.
When $r$ approaches infinite,
the oscillation amplitude near $\theta=0$ approaches a definite quantity and the curve has a definite wave shape.
This results from the asymptotic behavior of the differential cross section.
When $r\gg{d}$,
the expressions for the reduced electron flux $\overset{\sim}{j_r}$ reduces to a much simpler form
\begin{equation}
  \overset{\sim}{j_r}=\mathcal{A}_1^2|Y_{lm}(\beta_1,\varphi)|^2+\mathcal{A}_2^2|Y_{lm}(\beta_2,\varphi)|^2+2\mathcal{A}_1\mathcal{A}_2Y_{lm}(\beta_1,\varphi)Y_{lm}^*(\beta_2,\varphi)\cos(\Delta{S}-\frac{\pi}{2}),
\end{equation}
where
\begin{equation}
  \Delta{S}=kd\left(\cos\beta_2-\cos\beta_1+\frac{1}{\xi\eta}\ln\frac{1+\gamma\cos\beta_2}{1+\gamma\cos\beta_1}\right)
\end{equation}
and
\begin{equation*}
  \mathcal{A}_\nu=\left|\frac{\sin\beta_\nu}{\sin\theta\left(\frac{\partial\theta}{\partial\beta_\nu}\right)_r}\right|^{1/2}
\end{equation*}
with
\begin{equation}
  \left(\frac{\partial\theta}{\partial\beta_{\nu}}\right)_{r\to\infty}=1+\frac{1}{\xi\eta}\times\frac{\gamma}{1+\gamma\cos\beta_\nu}.
\end{equation}
For the middle part between the region near $\theta=0$ and that near $\theta=\pi$,
the variation is relatively weaker.
There is another point needing to be noticed in Fig. 4.
The differential cross section is infinite when $\theta\to\pi$ and thus it diverges at $\theta=\pi$.
However, it is integrable.
This will be discussed in the next section where the total photodetachment cross section is studied.
For $p_z$-wave phtodetachment, the similar situation exists, which is displayed in Fig. 5.

\subsubsection{The effect of the scaled energy}

Fig. $6$ shows the effect of the scaled energy on the differential cross section for $s$-wave photodetachment.
The locations of ${\theta}_c$ have been marked using dashed lines.
The special detection angle ${\theta}_c$ corresponds to electron emergence angle $\beta={\pi}/2$
  and has been defined in Sec. \rm{\uppercase\expandafter{\romannumeral2}}.
Increasing the scaled energy $\overset{\sim}E$,
the number of interferential peaks (the oscillation frequency) increases and the oscillation amplitude decreases.
In addition, the detection angle ${\theta}_c$ approaches  $\arccos{(d/r)}$.
That is to say, the affection of the central force is weaker as the scaled energy $\overset{\sim}E$ becomes larger.
For $p_z$-wave photodetachment,
there are analogous results which can be seen in Fig. $7$.

\subsubsection{The oscillation phase difference between $s$-wave and $p_z$-wave photodetachment}

 To make a comparison between $s$-wave and  $p_z$-wave photodetachment,
 the differential cross sections at infinite are plotted together in each subgraph of Fig. 7.
 With $d$ and $\alpha$ fixed at $300a_0$ and $1$ respectively,
 Fig. (a), Fig. (b) and Fig. (c) have different scaled energies:
  $\overset{\sim}E=2$,  $\overset{\sim}E=4$ and  $\overset{\sim}E=8$.
  Increasing scaled energy,
  the oscillation frequency increases, the oscillation amplitude decreases
  and ${\theta}_c$ gets closer to $\pi/2$.
  Between the two types of the photodetachment,
  from $\theta=0$ to $\theta=\pi$ there is a phase difference $\pi$ in differential cross section initially,
  but after $\theta={\theta}_c$ there is absolutely no phase difference.
   When $r\gg{d}$, ${\theta}_c$ has the asymptotic form
 \begin{equation}
   \theta_c=\arccos\left(\frac{1}{1+2\overset{\sim}E}\right).
 \end{equation}
  The change of the phase difference originates from the symmetry difference between an $s$-wave and a $p$-wave.
  For a $p_z$-wave,
  there is a node in the amplitude of the wave function when the electron ejects out with the polar angle $\pi/2$,
  whereas the $s$-wave amplitude has a spherically symmetric structure.
  From each subgraph of Fig. $8$,
  one can see that a phase reversal of differential cross section appears at $\theta_c$ for the $p_z$-wave.
 This can be seen more clearly in Fig. $9$.
 So for $p_z$-wave photodetachment,
 the position of ${\theta}_c$ is exactly where the phase reversal happens.

\subsubsection{The effect of the distance between the source point and the force center }
 The parameter $d$ also plays a nonnegligible role,
 which can be seen from Eq. $(37)$ to Eq. $(39)$.
 Assume $r\gg{d}$,
 Fig. $10$ displays the effect of the distance between the source point and the force center for $s$-wave photodetachment.
 The increase of $d$ yields the increase of the oscillation frequency.
 However, the amplitude is not altered.
  A similar phenomena exists for $p_z$-wave photodetachment.
 It is shown in Fig. $11$.
 Moreover, $\theta_c$ is labeled for the interference patterns with different $d$.
 The angle $\theta_c$ at which the phase reverses remains unchanged when $d$ changes.
 The reason is reflected by Eq. $(40)$,
 where ${\theta}_c$ is only dependent on scaled energy $Ed/\alpha$.

\section{total photodetachment cross section}

The total cross section can be obtained using the closed-orbit theory \cite{Du and Delos}.
When the detached electron ejects almost towards the force center,
it can return to the source region.
Hence near $\beta=\pi$,
there is a family of orbits among which one can select a central orbit as the closed orbit.
The returning electron wave interferes with the outgoing electron wave.
The interference between them produces the oscillation structure in the total cross section.
For convenient calculation, we will use the spherical coordinates $(r_s,\theta_s,\varphi_s)$ relative to the source point.

\subsection{Field-induced modulation function and reduced cross section}
According to the closed-orbit theory \cite{Du and Delos},
for photoniozation or photodetachment in an external circumstance the total cross section
always includes two parts:
 the smooth background (the total cross section without external field) and the oscillatory part.
 It can be expressed as
\begin{equation}
  \sigma=\sigma_0+\sigma_r
\end{equation}
where
\begin{equation}
  \sigma_0=-\frac{4\pi{E}_{ph}}{c}\mathrm{Im}\langle{D}\psi_i|\psi_{\rm{out}}\rangle
\end{equation}
and
\begin{equation}
  \sigma_r=-\frac{4\pi{E}_{ph}}{c}\mathrm{Im}\langle{D}\psi_i|\psi_{\rm{ret}}\rangle.
\end{equation}
$\psi_{\rm{ret}}$ represents the returning electron wave function \cite{Du and Delos}.

The returning wave function $\psi_{\rm{ret}}$ can be constructed using the semiclassical method.
We choose the same spherical surface $\Gamma$ mentioned before (in Sec. {\uppercase\expandafter{\romannumeral3}).
The electron waves start from it and travel along classical trajectories.
For the returning wave,
the trajectory crosses the symmetry axis of the system.
Thus the corresponding Maslov index equals $1$.
According to the semiclassical approximation,
the returning wave function is given as
\begin{equation}
  \psi_{\rm{ret}}=\psi_{\rm{out}}(R,\pi,\phi)A\exp[\mathrm{i}(S-\frac{\pi}{2})],
\end{equation}
with
\begin{equation}
  A=\frac{\alpha{R}}{4Ed^2}
\end{equation}
and
\begin{eqnarray}
  {\nonumber}S&=&\frac{2\sqrt{2d\alpha}}{\sqrt{\overset{\sim}E-1}}[\sqrt{\overset{\sim}E({\overset{\sim}E}-1)}+\ln{(\sqrt{\overset{\sim}E-1)}+\sqrt{\overset{\sim}E})}]\\
  \overset{\sim}E&\ge&1 .
\end{eqnarray}
$A$ and $S$ can be obtained using Eq. $(23)$ and Eq. $(27)$.

In Ref. $[17]$,
several succinct and general formulas are derived and used for the calculation of the total photodetachment cross section near a repulsive center. They also apply for the system considered in this article.
The expressions are written as follows.
\begin{equation}
  \sigma_0=\frac{2\pi{k}E_{ph}}{c}|C(k)|^2
\end{equation}
\begin{equation}
  \sigma_r=-(-1)^l(2l+1)|C(k)|^2\frac{2\pi{A}E_{ph}}{cR}\cos(S)\delta_{m0}.
\end{equation}
Ref. $[17]$ has given complete details about the derivation of the two expressions above.
The advantage of them is that they can be applied to different wave sources.
To analyse effectively, the field-induced modulation function \cite{B. C. Yang2} is used:
\begin{equation}
  \sigma=\sigma_0\mathcal{H}^c.
\end{equation}
Combine Eq. $(46)$ Eq. $(47)$ and Eq. $(48)$,
the expression of $\mathcal{H}^c$ has the form
\begin{equation}
  \mathcal{H}^c=1-(-1)^l(2l+1)\frac{A}{kR}\cos(S)\delta_{m0}.
\end{equation}
 Further more, a reduced total cross section \cite{B. C. Yang2} is defined as
 \begin{equation}
   \overset{\sim}{\sigma}=k^{2l+1}\mathcal{H}^c.
 \end{equation}

In addition,
the total cross section can also be calculated by integrating the differential cross section obtained early.
The two results given using the two methods (using the closed-orbit theory and integrating the differential cross section)
coincide very well.
This can be seen in Fig. $13$ and Fig. $14$ which display the calculation results of the reduced cross section for
an $s$-wave source and a $p_z$-wave source respectively.

\subsection{Calculations for the field-induced modulation function and reduced total cross section}

We still consider the $s$-wave source and $p_z$-wave source.
In Fig. $12$,
the field-induced modulation functions are compared between the two types of wave source.
The oscillatory structures appear clearly.
The variation of phase and amplitude can be modulated through $d$ simultaneously.
Increasing $d$ results in the decrease of the oscillatory amplitude and the increase of the frequency,
which is reflected by Eq. $(46)$ and Eq. $(50)$.
From Fig. $12(a)$ to Fig. $12(c)$,
the phase difference $\pi$ always exists whichever $d$ is.
Analysis may be carried out using Eq.$(46)$.
$S$ depends totally on the scaled energy and the product of $d$ and $\alpha$.
It is irrelevant to the wave source.
However the factor $(-1)^l$ can alter the phase by $\pi$ when $l$ is changed by $1$.
In other words,
the parity of the spherical harmonic function $Y_{lm}$ can influence the phase.
Considering that $l=0$ for an $s$-wave and $l=1$ for a $p$-wave,
a phase difference $\pi$ arises in the filed-induced modulation functions for the two types of wave source.
This relationship coincides with the phase difference in the differential cross section between the two types of wave source.

The reduced total cross sections are plotted for an $s$-wave source and a $p_z$-wave source in Fig. $13$ and Fig. $14$ respectively.

\subsection{Photodetachment of $\rm{H^{-}}$ near an attractive center }

In the former discussion, the concrete physical model of the wave source is not given.
So in the expression of the outgoing wave function $\psi_{\rm{out}}$ (Eq. $(18)$),
the factor $C(k)$ is unknown.
During the numerical calculation, one has to eliminate it by virtue of a reduced electron flux,
a field-induced modulation function, etc.

Now, we consider the photodetachment of $\rm{H^-}$ near an attractive force center with the positive charge number $\alpha$.
The distance between the two point is still $d$.
The polarization direction is along the $z$ axis which joins the $\rm{H^-}$ ion and the force center.
After detachment, the system produces the outgoing $p_z$-wave.
To construct the returning wave function,
the selected spherical surface $\Gamma$ mentioned earlier is used.
The expression of $\psi_{\rm{out}}$ can be written as
\begin{equation}
  \psi_{\rm{out}}=\frac{4Bk\rm{i}}{(k_b^2+k^2)^2}\frac{\exp(\rm{i}kR)}{R}\cos\beta.
\end{equation}
$B$ is a normalization constant and equals $0.31552$ \cite{Du1} and $k_b=\sqrt{2E_b}$.
$E_b$ represents the binding energy and is about $0.7542\,\rm{eV}$ for $\rm{H^-}$.
Combining the above equation with Eq. $(17)$,
the factor $C(k)$ reads
\begin{equation}
  C(k)=\frac{\rm{i}kB}{N_{10}E_{ph}^2}.
\end{equation}
Substituting Eq. $(53)$ into Eq. $(47)$ gives
\begin{equation}
  \sigma_0=\frac{16\pi^{2}\sqrt{2}B^{2}E^{3/2}}{3c{(E_b+E)}^3},
\end{equation}
which is consitstent  with the result obtained in previous literatures \cite{Rau, Du3}.
Substituting Eq. $(53)$ into Eq. $(48)$ leads to the oscillatory term
\begin{equation}
  \sigma_r=\frac{4\pi^{2}B^{2}}{c{(E_b+E)}^3}\frac{\alpha}{d^{2}}\cos{(S)}.
\end{equation}
According to Eq. $(41)$,
the total cross section is the sum of the above two terms
\begin{equation}
  \sigma=\frac{16\pi^{2}\sqrt{2}B^{2}E^{3/2}}{3c{(E_b+E)}^3}+\frac{4\pi^{2}B^{2}}{c{(E_b+E)}^3}\frac{\alpha}{d^{2}}\cos{(S)}.
\end{equation}

Fig. $15$ shows the calculation results of the total photodetachment cross section for $\rm{H^-}$ near an attractive force center.
We set $d=200\,a_0$ and $\alpha=1$.
The oscillatory structure can be observed in the figure.
It comes from the interference between the outgoing and returning electron wave.

\subsection{Comparing photodetachment near an attractive center with that near a repulsive center and that in a uniform electric field}
\subsubsection{Comparison based on the photodeatachment of ${\rm{H}}^-$}
In Ref. $[17]$,
the total phtodetachment cross section of $\rm{H^-}$ near a repulsive center has been compared with that in a uniform field whose strength $F=\alpha/{d^2}$.
$d$ and $\alpha$ have the same meanings mentioned before.
Here we compare the three conditions together to find out their similarities and differences.

The three physical system models are set as follows:
$\rm{H^-}$ near a force center with the positive charge number $\alpha$,
$\rm{H^-}$ near a force center with the negative charge number $\alpha$
and $\rm{H^-}$ in a uniform static electric field.
The first two systems have similar configurations: the repulsive center locates where the attractive center is.
 In the third system, the field strength $F$ is $\alpha/{d^2}$.
The direction is
parallel to the symmetry axis of the first two systems and points from the source point to the force center.
The main purpose of the above configurations is to compare and analyse more conveniently.

The comparison results are displayed in Fig. $16$, in which $d=200\,a_0$, $\alpha=1$ and $F=\alpha/{d^2}=128.55\,\rm{kV/cm}$.
All the total photodetachment cross sections have oscillatory structures.
This reflects that all of them have closed orbits, which can be deduced from the trajectories of detached electrons.
In each system, the electron ejecting towards the force center (along the direction of the electric field for the third one)
can return to the region of the source point, and thus the only closed orbit forms.
The oscillation amplitudes for the three systems are exactly the same.
The increase of photon energy $E_{ph}$ yields the decrease of the amplitude.
However their oscillation frequencies are different.
For the uniform field, the oscillation frequency is much higher than the others.
Compare the other two: the oscillation frequency for the repulsive central field and that for the attractive central field,
one can find that the former is a little higher than the latter.
For the photdetachment of $\rm{H^-}$ in the uniform field,
the oscillation frequency increases when the photon energy $E_{ph}$ increases.
In the case of the other two situations, the frequecncy decreases when $E_{ph}$ increases.

\subsubsection{The generalization and explanation of the relationships}

Fig. $16$ displays the total photodetachment of ${\rm{H}}^-$ in the three types of external field.
These comparison results can been seen more clearly in the field-induced modulation functions.
Fig. $17$ displays the corresponding modulation functions of the three systems.
However, these relationships are not only confined to the photodetachment of $\rm{H}^-$.
They also apply for the photodetachment of other anions.

To understand this generalization, one may go back to Eq. $(49)$ which is general for photodetachment in imposed
external fields.
For the photodetachment of $\rm{H}^-$ in the three environments,
the smooth background is invariable and can be written as Eq. $(54)$.
For the photodetachment of an arbitrary single electron anion in the three environments,
the smooth background is also a common factor and has been expressed in the form of Eq. $(47)$.
The difference comes from the field-induced modulation function
\begin{equation}
  \mathcal{H}^F=1-\frac{3A}{kR}\cos{S}.
\end{equation}
Different external fields give different modulation functions.
Here, $A$ and $S$ are the main elements considered in $\mathcal{H}^F$.

For the repulsive central force field, the amplitude expression Eq. $(45)$  also applies\cite{B. C. Yang2}.
As for the uniform electric field \cite{Du6},
\begin{equation}
  A=\frac{RF}{4E}.
\end{equation}
Substitute $F=\alpha/{d^2}$ to the above equation,
the above expression becomes identical with Eq. $(45)$.
So the amplitudes are identical for the three systems.
In addition, the equivalence of $A$ can be understood through the dynamics of the three systems.
Fig. $16$ depicts a family of trajectories with emergence angle $\beta$ near $\pi$ for each system.
The central trajectories correspond to the closed orbits.
We distinguish the trajectories for different systems using different colors.
The trajectories with the same $\beta$ for the three systems cross each other when go back to the source region.
Linking the intersections gives the dash line in Fig. $16$.
In real space, the intersections constitute a curve,
 on which the flux densities of the three conditions are equal for the fluxes whose directions are along the trajectories.
 Along the $z$ axis,
 the three trajectories coincide
 and thus the returning waves have equal probability densities.
 This causes the equivalence of amplitude $A$ observed in the spectra.

The classical action along the closed orbit of the electron in the attractive central force field is given in Eq. $(46)$.
For the system with the repulsive force center \cite{B. C. Yang2},
\begin{equation}
  S_{\rm{repulsive}}=\frac{2\sqrt{2d\alpha}}{\sqrt{1+\overset{\sim}E}}[\sqrt{\overset{\sim}{E}(1+{\overset{\sim}E})}-\ln{(\sqrt{\overset{\sim}E+1}+\sqrt{\overset{\sim}E})}].
\end{equation}
For that with the uniform electric field,
\begin{equation}
  S_{\rm{uniform}}=\frac{4\sqrt{2da}}{3}{\overset{\sim}E}^{3/2}.
\end{equation}
In Fig. $17$,
the classical actions and their differentials to energy
are depicted.
The variation  of ${\rm{d}}S/{\rm{d}}E$ determines the variation of oscillation frequency in the spectra.
In Fig. $17(b)$,
 ${\rm{d}}S/{\rm{d}}E$ for the system with the uniform field is much larger than the other two.
Increasing the photon energy $E_{ph}$ causes the increase of ${\rm{d}}S/{\rm{d}}E$.
${\rm{d}}S/{\rm{d}}E$ for the attractive central force field is a little smaller than that for the repulsive central force field.
Both of them decrease when $E_{ph}$ increases.
These relationships and variations are directly reflected
in the aspect of the oscillation frequency of spectra (Fig. $16$ and Fig. $17$).

\section{conclusions}

In this article, we have studied the photodetachment of a single electron anion near an attractive force center.
After the analysis of the orbit equation and classical motion,
the classically propagating wave function is constructed and the detached electron flux obtained.
Two types of wave source are considered: the $s$-wave source and the $p_z$-wave source.
A spherical detector centered at the force center is assumed.
There are oscillations in the detachedelectron angular distribution pattern.
The detector radius, the scaled energy , the distance $d$ etc can affect the oscillatory structure.
For a $p_z$-wave photodetachment, a phase reversal always occurs at the detection angle ${\theta}_c$ which corresponds to the emergence angle
$\beta=\pi/2$ .
This leads to the change of the phase difference of differential cross section at infinite between the $s$-and $p_z$- wave photodetachment.
Using the closed-orbit theory,
the total photodetachment cross sections are calculated.
Also, the differential cross section obtained earlier can be integrated to get the total cross sections.
Numerical calculations show that the two methods are in agreement with each other.

We use the field-induced modulation function, through which one can analyse some properties more easily and directly.
The photodetachment of $\rm{H^-}$ near an attractive force center is also studied and
the total cross sections are computed.
Several comparisons among the photodeatachments near an attractive force center, near a repulsive force center and in a uniform electric field are made.
One can find the similarities of their amplitudes and the differences of their phases.

\begin{center}
{\bf ACKNOWLEDGMENTS}
\end{center}
\vskip8pt This work was supported by NSFC grant No. 11074260 and No.
11121403.

 \begin{appendix}
 \section{The derivation of eq. (24)}
 Define
  \begin{equation}
   {\theta}_b=\arccos\left(\frac{\frac{d}{r}{\xi}^2\sin^2\beta-1}{\varepsilon}\right).
 \end{equation}
Use this definition,
Eq. $(4)$ becomes
\begin{equation}
  \theta={\theta}_0+{\theta}_b.
\end{equation}
${\theta}_0$ is the perihelion angle.
\begin{equation}
  \cos{\theta}_0=\frac{\xi^2\sin^2\beta-1}{\varepsilon},
\end{equation}
and
\begin{equation}
  \sin{\theta}_0=\frac{-\xi^2\sin\beta\cos\beta}{\varepsilon}.
\end{equation}
 The above two equations yield
 \begin{equation}
   \left(\frac{\partial \cos{\theta}_0}{\partial\beta}\right)_r=-\sin{\theta}_0\left(1+\frac{1+\eta^2}{\varepsilon^2}\right).
 \end{equation}
 Then
 \begin{equation}
   \left(\frac{\partial {\theta}_0}{\partial \beta}\right)_r=1+\frac{1+\eta^2}{\varepsilon^2}.
 \end{equation}
 Similarly,
 \begin{equation}
   \left(\frac{\partial {\theta}_b}{\partial \beta}\right)_r=-\frac{\eta^2+\frac{d}{r}(1+\varepsilon^2)}{\varepsilon^2}\times\frac{k}{p_r}\cos\beta.
 \end{equation}
 So
 \begin{equation}
   \left(\frac{\partial {\theta}}{\partial \beta}\right)_r=1+\frac{1+\eta^2}{\varepsilon^2}-\frac{\eta^2+\frac{d}{r}(1+\varepsilon^2)}{\varepsilon^2}\times\frac{k}{p_r}\cos\beta.
 \end{equation}

 \section{The derivation of eq. (28)}
\begin{equation}
  S_d=\int^{{\beta}_\mu}_0kd\sin\beta\left(\frac{\partial\theta}{\partial\beta}\right)_r{\rm{d}}\beta
\end{equation}
Substitute Eq. (A.8) to the above expression,
\begin{equation}
  S_d=kd\times\left[(1-\cos{\beta}_\nu)+(1+\eta^2)\Gamma-\xi(\eta^2+\frac{d}{r})\Omega-\xi\times\frac{d}{r}\times\Xi\right].
\end{equation}
\begin{equation}
  \Gamma=\int^{\beta_\nu}_0\frac{\sin\beta}{1+\xi^2\eta^2\sin^2\beta}{\rm{d}}\beta.
\end{equation}
\begin{equation}
  \Omega=\int^{\beta_\nu}_0\frac{\sin\beta\cos\beta}{(1+\xi^2\eta^2\sin^2\beta)\times\sqrt{\eta^2+2\frac{d}{r}-\frac{d^2}{r^2}\xi^2\sin^2\beta}}{\rm{d}}\beta.
\end{equation}
\begin{equation}
  \Xi=\int^{\beta_\nu}_0\frac{\sin\beta\cos\beta}{\sqrt{\eta^2+2\frac{d}{r}-\frac{d^2}{r^2}\xi^2\sin^2\beta}}{\rm{d}}\beta.
\end{equation}
After integration,
one gets
\begin{equation}
  \Gamma=\frac{1}{\xi^2-1}\times\frac{1}{\xi\eta}\left(\ln\varepsilon+\ln\frac{1+\gamma}{1+\gamma\cos\beta_\nu}\right),
\end{equation}
\begin{equation}
  \Omega=\frac{1}{\xi^2\eta(\eta^2+\frac{d}{r})}\times\left(\ln\varepsilon+\ln\frac{\eta^2+\frac{d}{r}+\frac{\xi\eta}{k}p_{0r}}{\eta^2+\frac{d}{r}+\frac{\xi\eta}{k}p_{{\nu}r}}\right),
\end{equation}
and
\begin{equation}
   \Xi=\frac{k}{\xi}\times\frac{\sin^2\beta}{p_{0r}+p_{{\nu}r}}.
\end{equation}

Substitute Eq. (B6) Eq. (B7) and Eq. (B8) to Eq. (B2), we obtain Eq. $(28)$.

 \end{appendix}


\newpage

\begin{figure}
  \centering
  \includegraphics[scale=0.8]{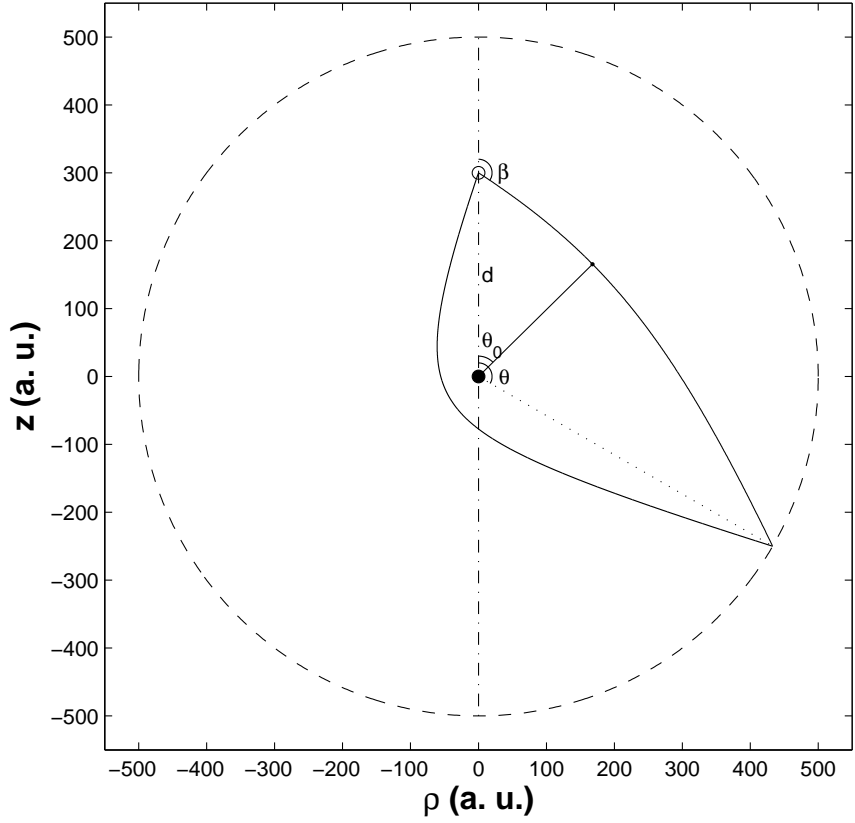}\\
  \caption{The theoretical model for the photodetachment of an anion near an attractive center.
  The open and solid circles represent the single electron anion and force center respectively.
  The force center has $\alpha$ positive charges.
  The dash-dot line represents the symmetry axis of the system.
  The dashed circle denotes the assumed spherical detector centered at the force center.
  $\beta$ is the emergence angle of the detached electron,
  $\theta$ the detection angle
  and ${\theta}_0$ the perihelion angle.
  $d$ is the distance between the anion and force center.
  $d=300\,a_0$.
  The detection distance is $500\,a_0$.
  The scaled energy $\overset{\sim}E=Ed/\alpha=2$.
  $a_0$ represents the Bohr radius.}
\end{figure}

\begin{figure}
  \centering
  \includegraphics[scale=0.8]{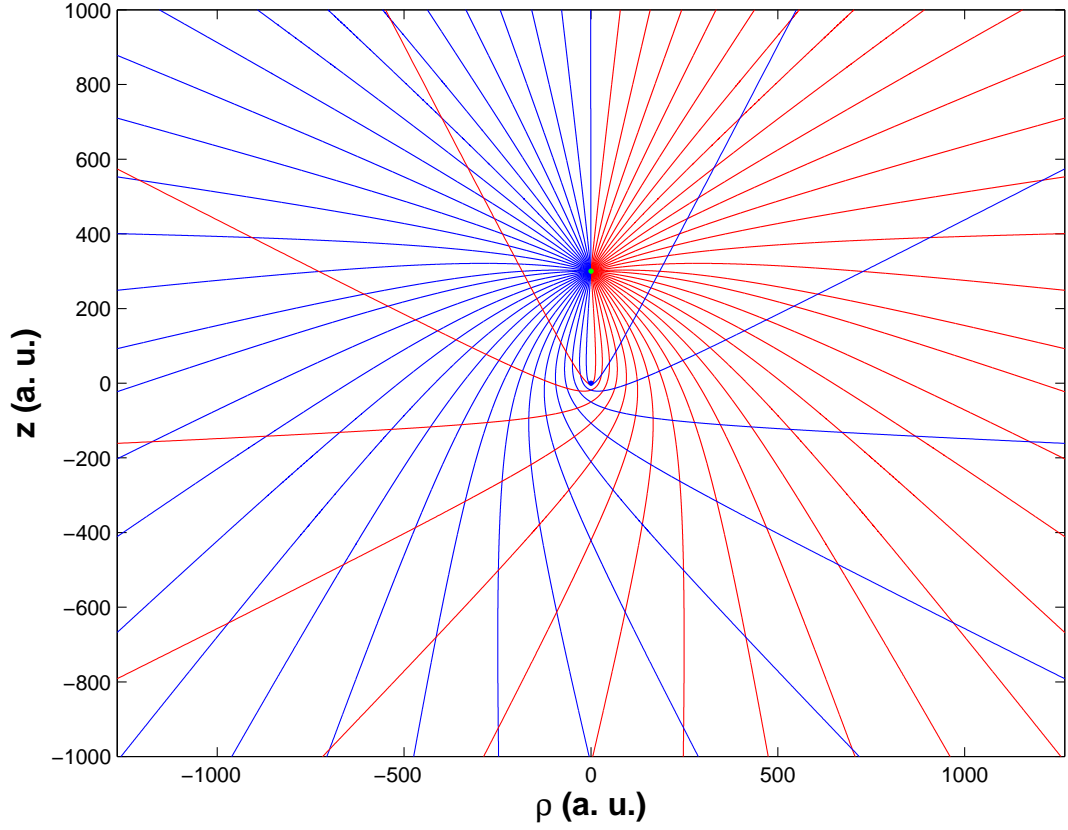}\\
  \caption{Some electron trajectories with different emergence angles.
  The green dot denotes the source point.
  The blue dot is the force center with the positive charge number $\alpha=1$.
   The distance between the source and force center $d=300\,a_0$ and the scaled energy $\overset{\sim}E=Ed/\alpha=2$.
   $a_0$ is the Bohr radius.}
\end{figure}

\begin{figure}
  \centering
  \includegraphics[scale=0.8]{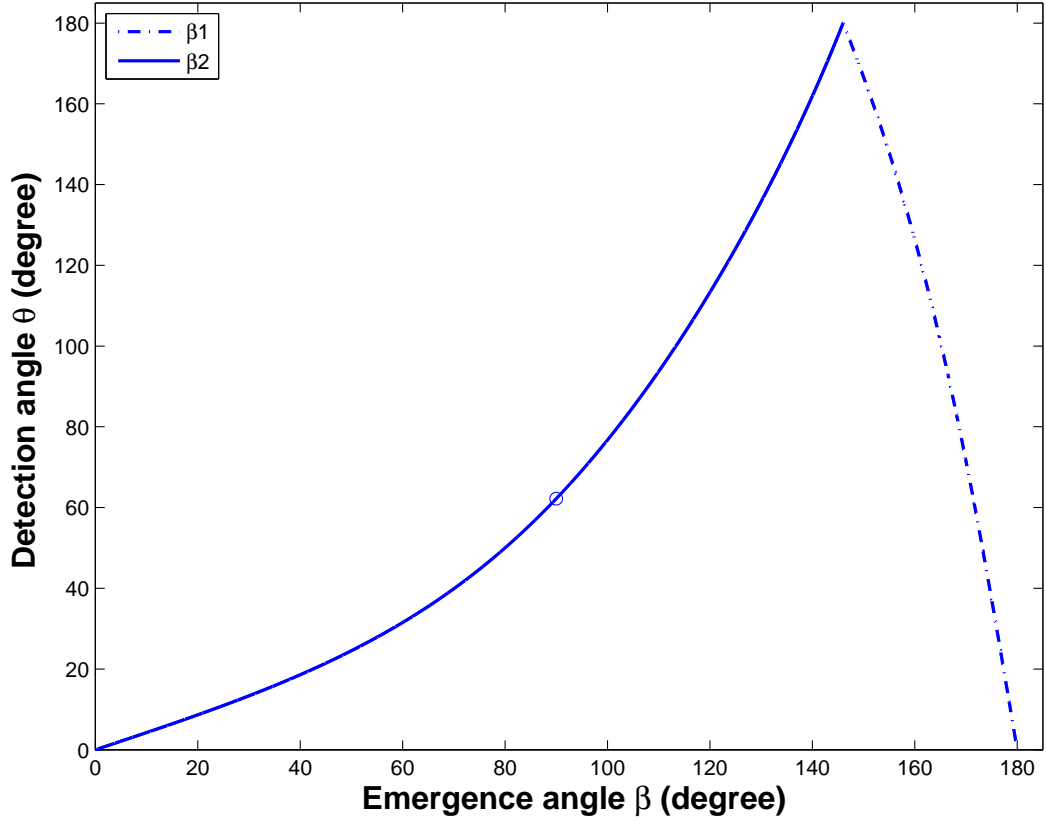}\\
  \caption{The detection angle $\theta$ vs the emergence angle $\beta$.
  Detector radius $r=500\,a_0$, the distance between the source and force center $d=300\,a_0$, and the scaled energy $\overset{\sim}E=Ed/\alpha=2$.
  $a_0$ is the Bohr radius.
  The circle marks the position of $\theta_c$ which corresponds to the emergence angle $\pi/2$.}
\end{figure}

\begin{figure}
\includegraphics[scale=.80,angle=-0]{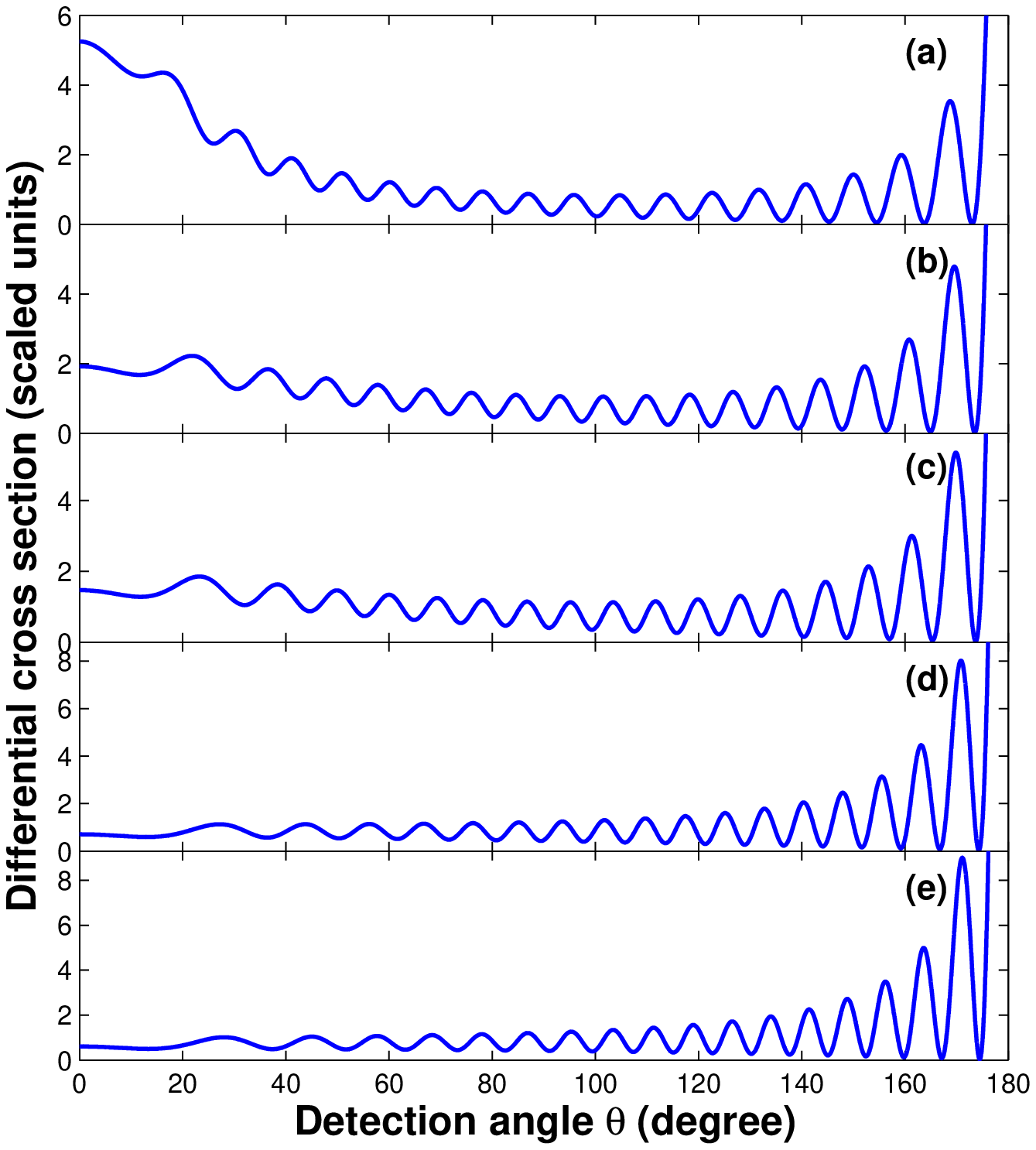}
\caption{Differential cross section as the function of detection angle $\theta$ for $s$-wave detachment.
   $\alpha=1$, $d=300\,a_0$ and the scaled energy $\overset{\sim}E=Ed/\alpha=2$.
  (a) $r=500\,a_0$, (b) $r=800\,a_0$, (c) $r=1000\,a_0$, (d) $r=5000\,a_0$, (e) $r\gg{d}$.
  $a_0$ is the Bohr radius.
  The variation of detection distance $r$ has a significant influence on the differential cross section near $\theta=0$ and $\theta=\pi$.}
\end{figure}

\begin{figure}
\includegraphics[scale=.80,angle=-0]{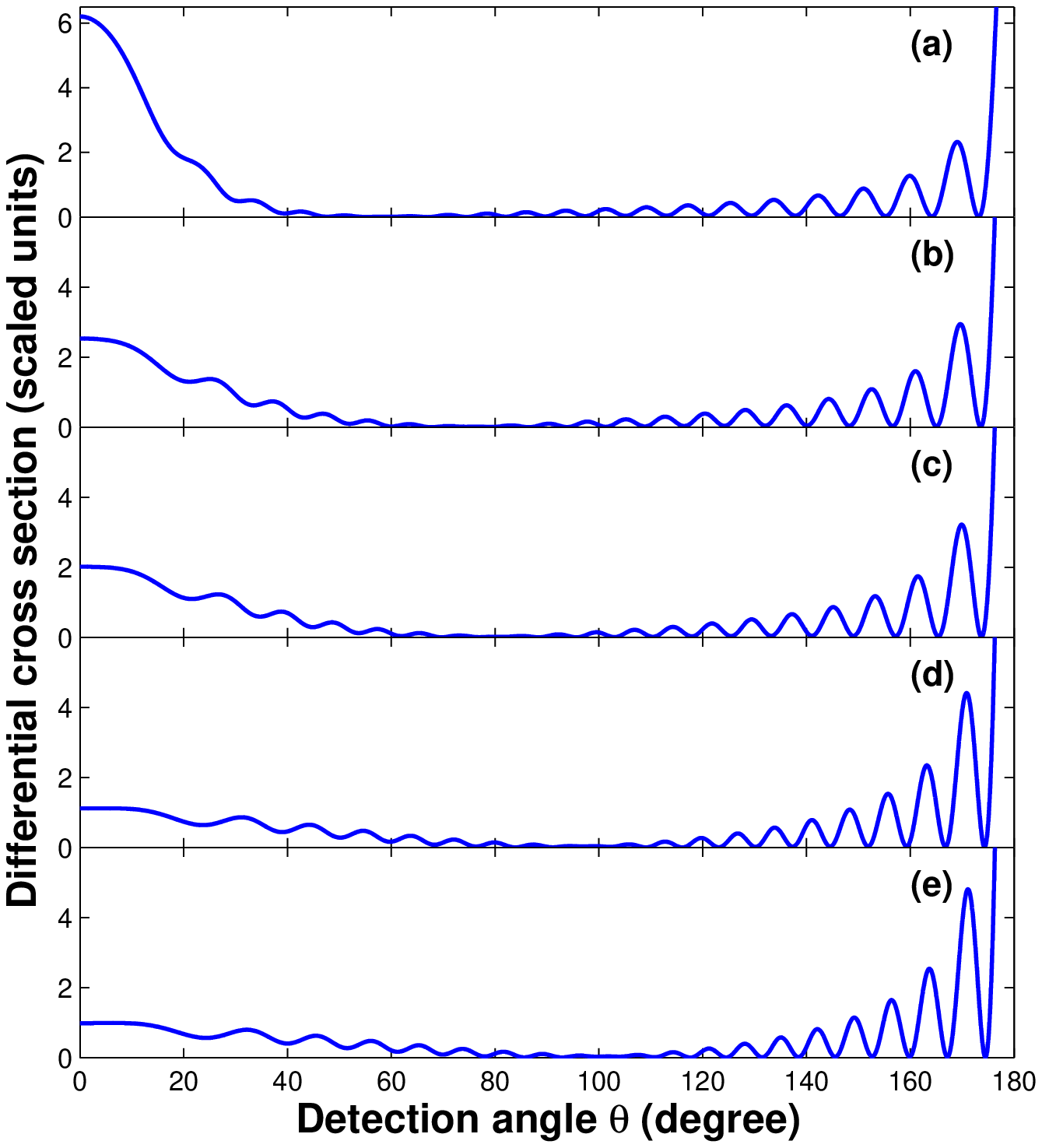}
\caption{Differential cross section as the function of detection angle $\theta$ for $p$-wave detachment with $z$ linear polarization.
   $\alpha=1$, $d=300\,a_0$ and the scaled energy $\overset{\sim}E=Ed/\alpha=2$.
  (a) $r=500\,a_0$, (b) $r=800\,a_0$, (c) $r=1000\,a_0$, (d) $r=5000\,a_0$, (e) $r\gg{d}$.
  $a_0$ is the Bohr radius.
  The variation of detection distance $r$ has a significant influence on the differential cross section near $\theta=0$ and $\theta=\pi$.}
\end{figure}

\begin{figure}
  \centering
  \includegraphics[scale=0.8]{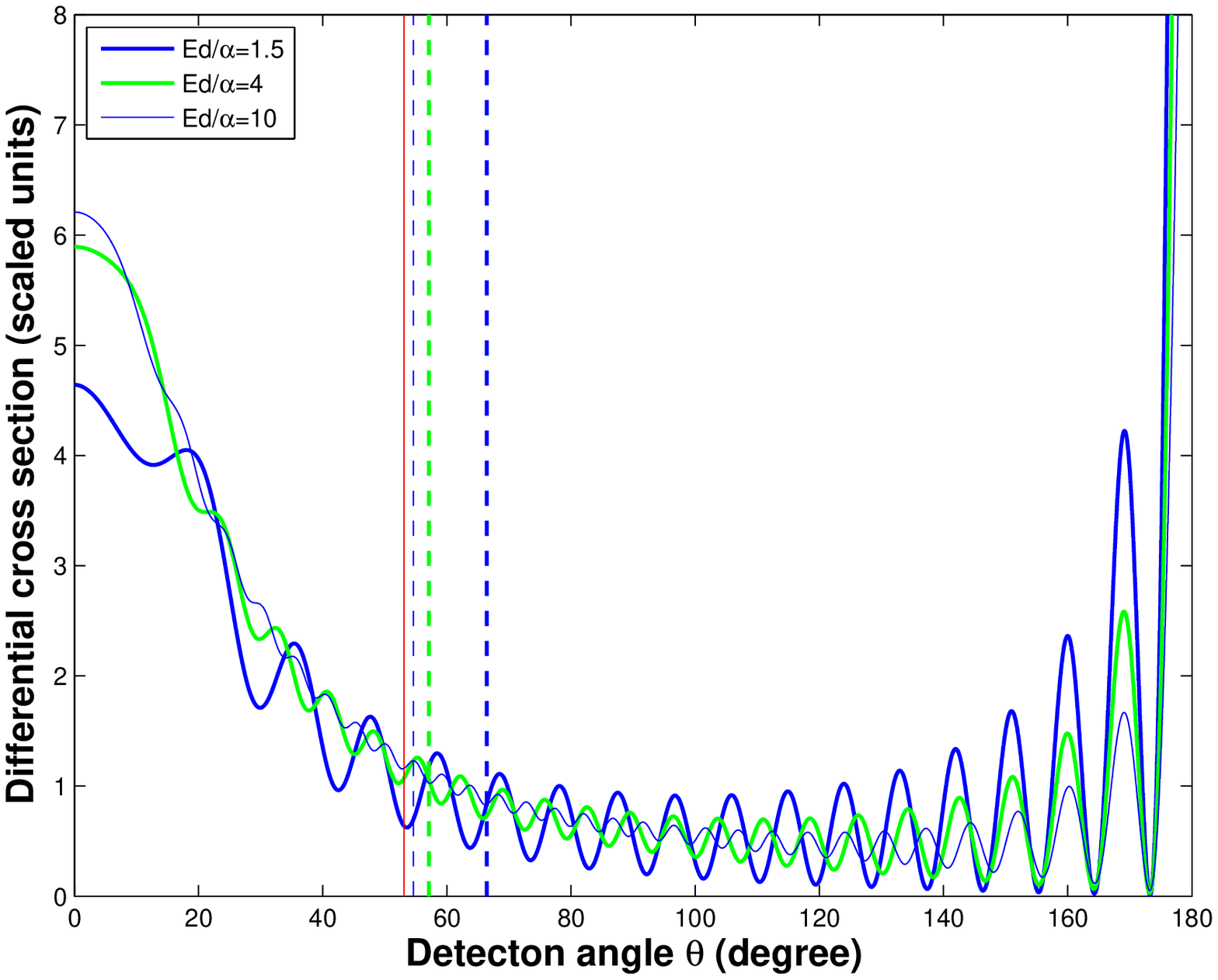}\\
  \caption{Differential cross section as the function of detection angle $\theta$ for $s$-wave photodetachment.
   $\alpha=1$, $d=300\,a_0$ and $r=500\,a_0$.
   $a_0$ is the Bohr radius.
   The locations of the corresponding ${\theta}_c$ are all marked:
   the blue thick dashed line for $\overset{\sim}E=1.5$,
   the green thick dashed line for $\overset{\sim}E=4$,
   and the blue thin dashed line for $\overset{\sim}E=10$.
   The red thin solid line represents the position of $\theta=\arccos{(d/r)}=53.13$ (degree).
   When the scaled energy $\overset{\sim}E=Ed/\alpha$ is increased,
   the interferential peak number increases but the amplitude decreases.}
\end{figure}

\begin{figure}
  \centering
  \includegraphics[scale=0.8]{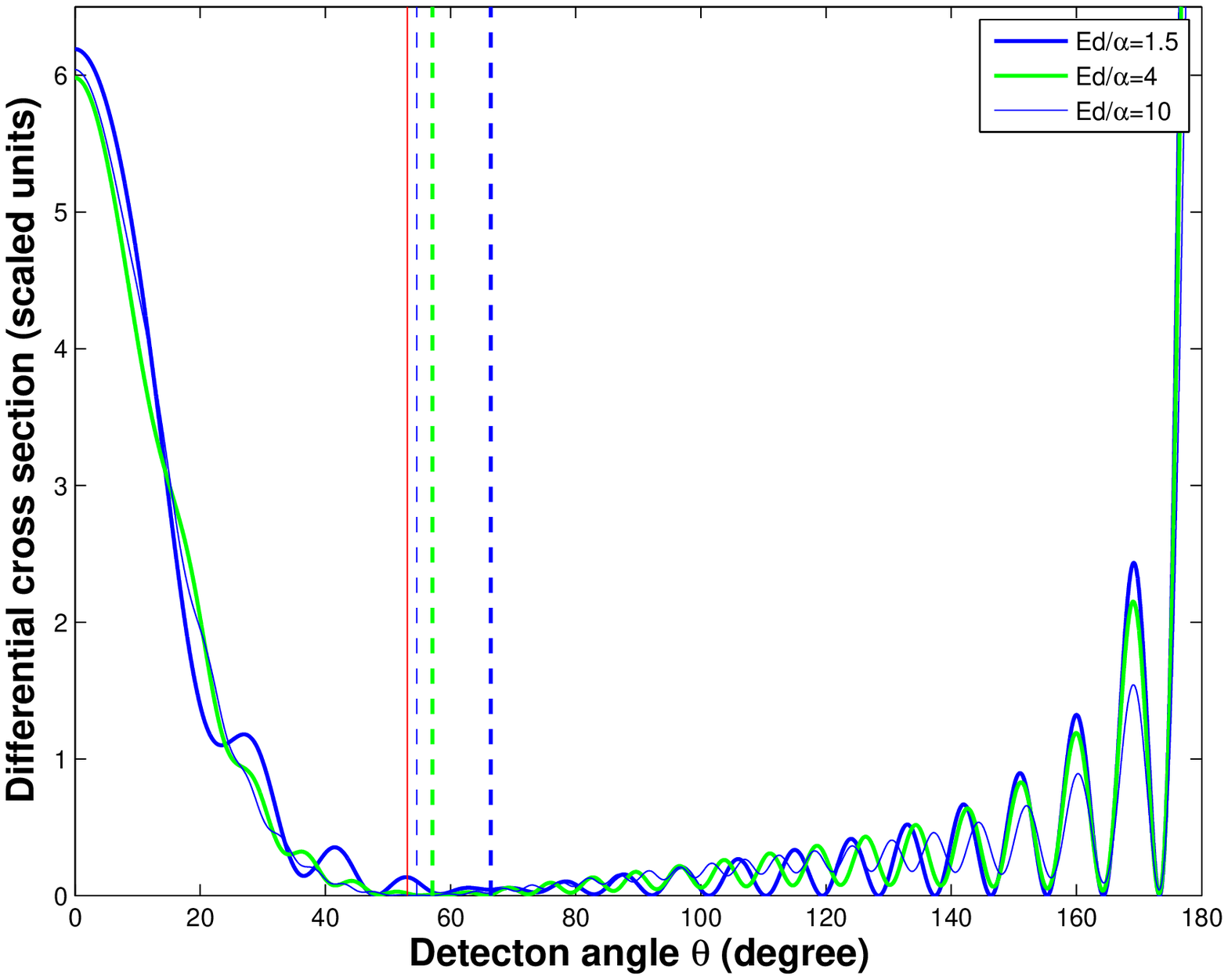}\\
  \caption{Differential cross section as the function of detection angle $\theta$ for $p_z$-wave photodetachment.
   $\alpha=1$, $d=300\,a_0$ and $r=500\,a_0$.
   $a_0$ is the Bohr radius.
   The locations of the corresponding ${\theta}_c$ are all marked:
   the blue thick dashed line for $\overset{\sim}E=1.5$,
   the green thick dashed line for $\overset{\sim}E=4$,
   and the blue thin dashed line for $\overset{\sim}E=10$.
   The red thin solid line represents the position of $\theta=\arccos{(d/r)}=53.13$ (degree).
   When the scaled energy $\overset{\sim}E=Ed/\alpha$ is increased,
   the interferential peak number increases but the amplitude decreases.}
\end{figure}

\begin{figure}
  \centering
  \includegraphics[scale=.80]{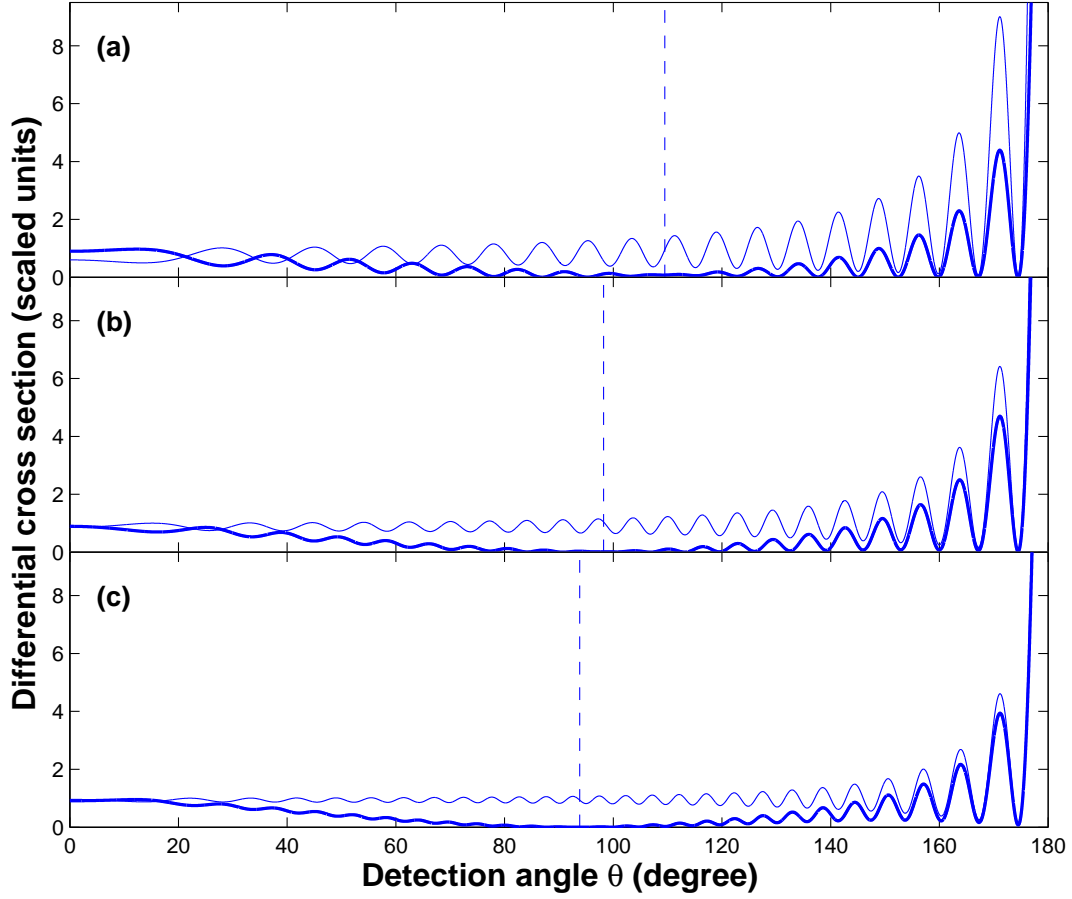}\\
  \caption{The differential cross section as the function of detection angle $\theta$
  at infinitely large detector radius $r\gg{d}$.
   $\alpha=1$, $d=300\,a_0$.
   $a_0$ is the Bohr radius.
   (a) Scaled energy $\overset{\sim}E=Ed/\alpha=2$, (b) $\overset{\sim}E=4$, (c) $\overset{\sim}E=8$.
  The thin line represents $s$-wave detachment.
  The thick line represents $p$-wave detachment with $z$ linear polarization
  The dashed line indicates the position of ${\theta}_c$.
  For the two types of detachment,
  before ${\theta}_c$ there is a phase difference $\pi$ and
  after ${\theta}_c$ there is no phase difference.}
\end{figure}

\begin{figure}
  \centering
  \includegraphics[scale=.8]{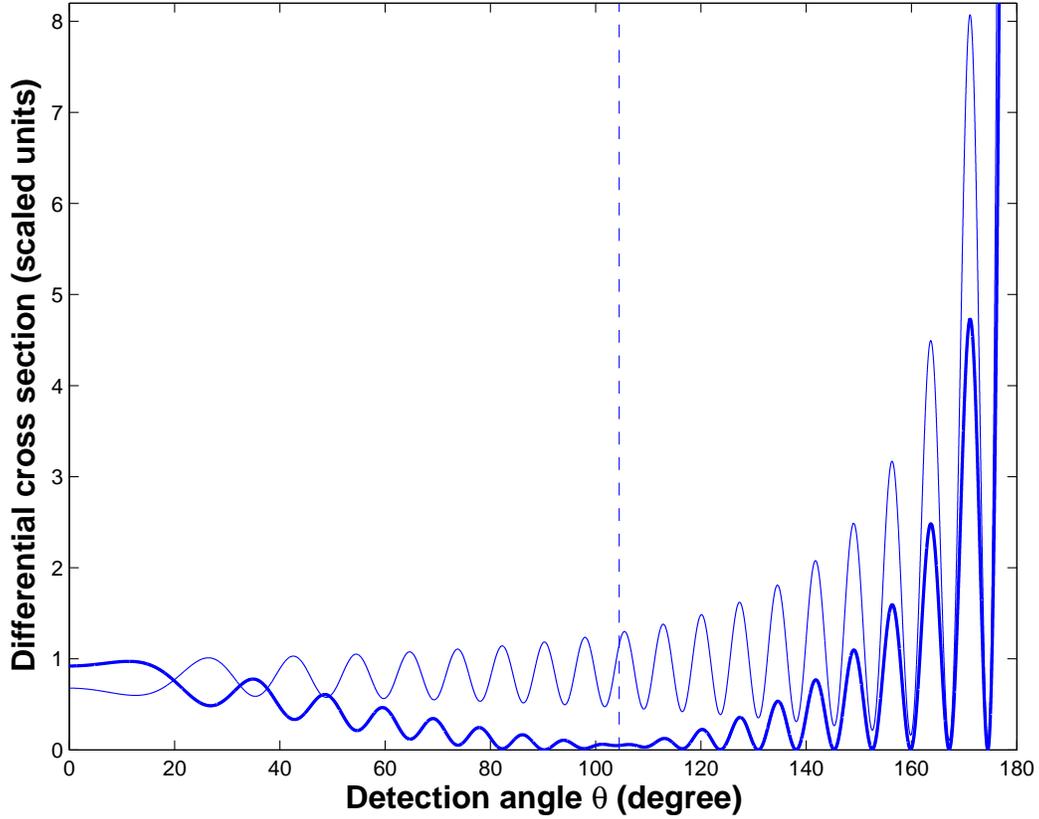}\\
  \caption{The differential cross section as the function of detection angle $\theta$
  at infinitely large detector radius $r\gg{d}$.
   $\alpha=1$, $d=300\,a_0$ and the scaled energy $\overset{\sim}E=Ed/\alpha$ is $2$.
  The thin line represents $s$-wave detachment.
  The thick line represents $p$-wave detachment with $z$ linear polarization
  The dashed line indicates the position of ${\theta}_c$.
  For the two types of detachment,
  before ${\theta}_c$ there is a phase difference $\pi$ and
  after ${\theta}_c$ there is no phase difference.
   }
\end{figure}

\begin{figure}
  \centering
  \includegraphics[scale=.8]{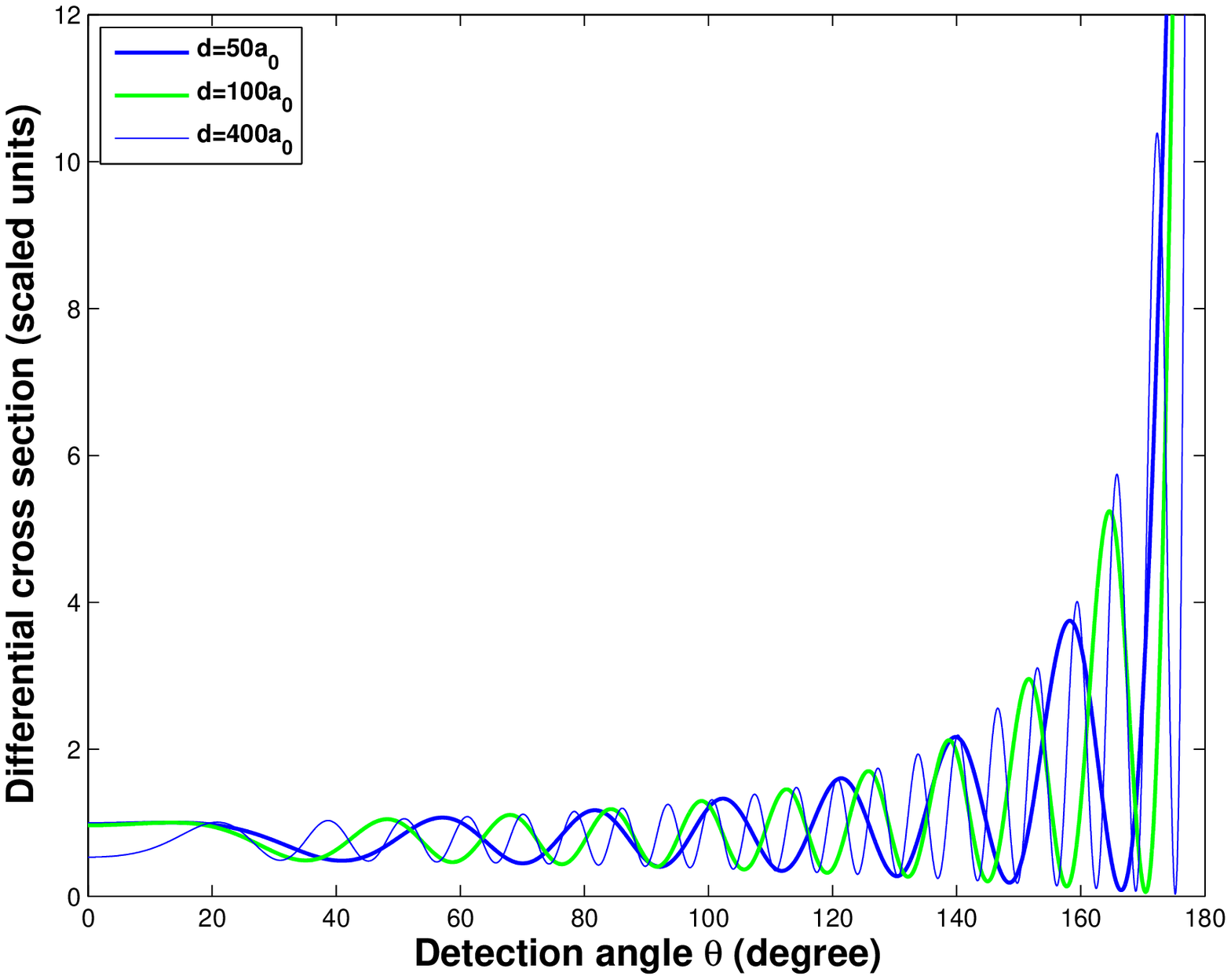}\\
  \caption{The differential cross section as the function of detection angle $\theta$
  at infinitely large detector radius $r\gg{d}$ for an $s$-wave source.
   $\alpha=1$ and the scaled energy $\overset{\sim}E=Ed/\alpha$ is $1.5$.
   When $d$ is increased, the interferential peak number increases but the amplitude doesn't change. }
\end{figure}

\begin{figure}
  \centering
  \includegraphics[scale=.8]{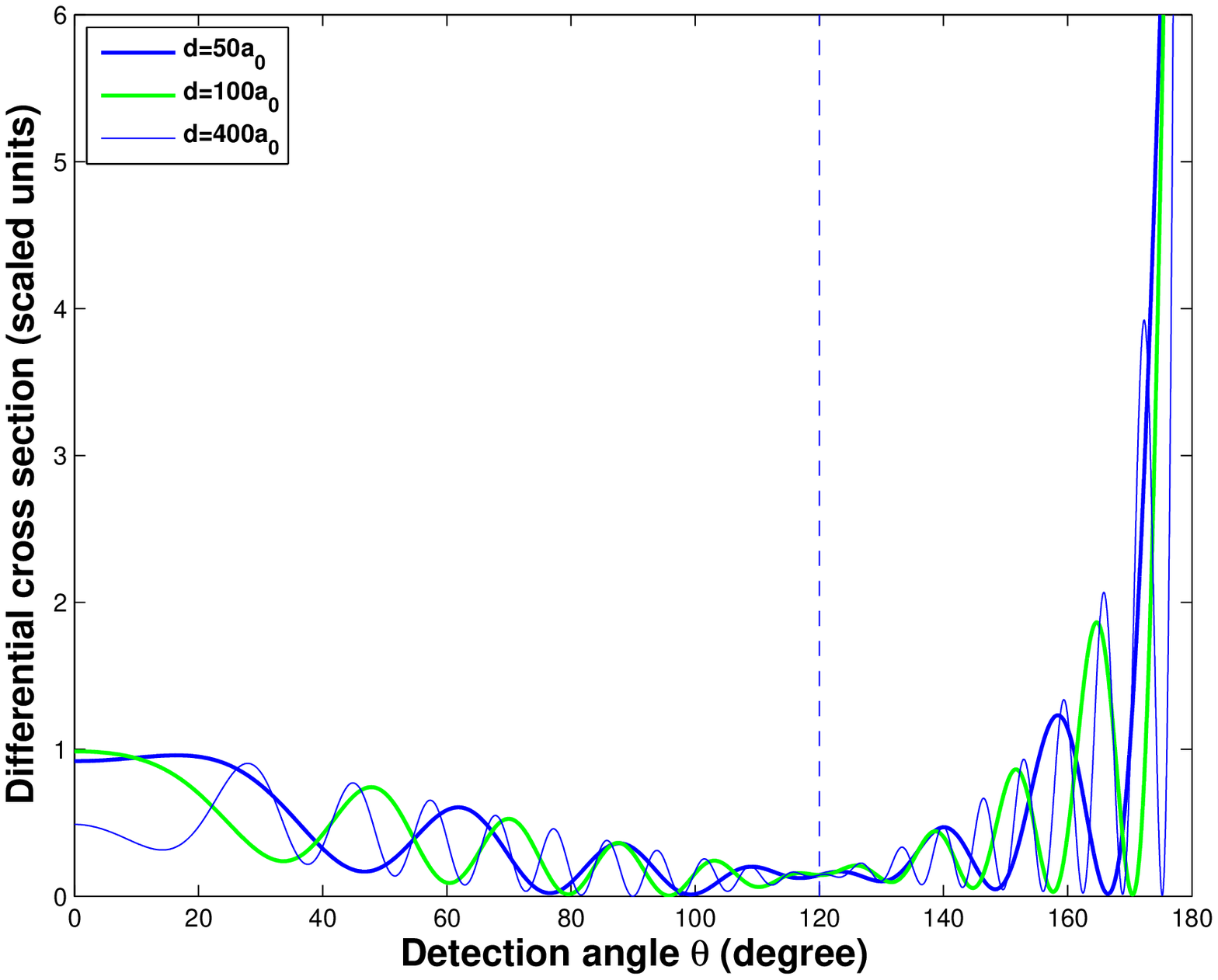}\\
  \caption{The differential cross section as the function of detection angle $\theta$
  at infinitely large detector radius $r\gg{d}$ for a $p_z$-wave source.
   $\alpha=1$ and the scaled energy $\overset{\sim}E=Ed/\alpha$ is $1.5$.
   When $d$ is increased, the interferential peak number increases but the amplitude doesn't change.}
\end{figure}

\begin{figure}
  \centering
  \includegraphics[scale=.8]{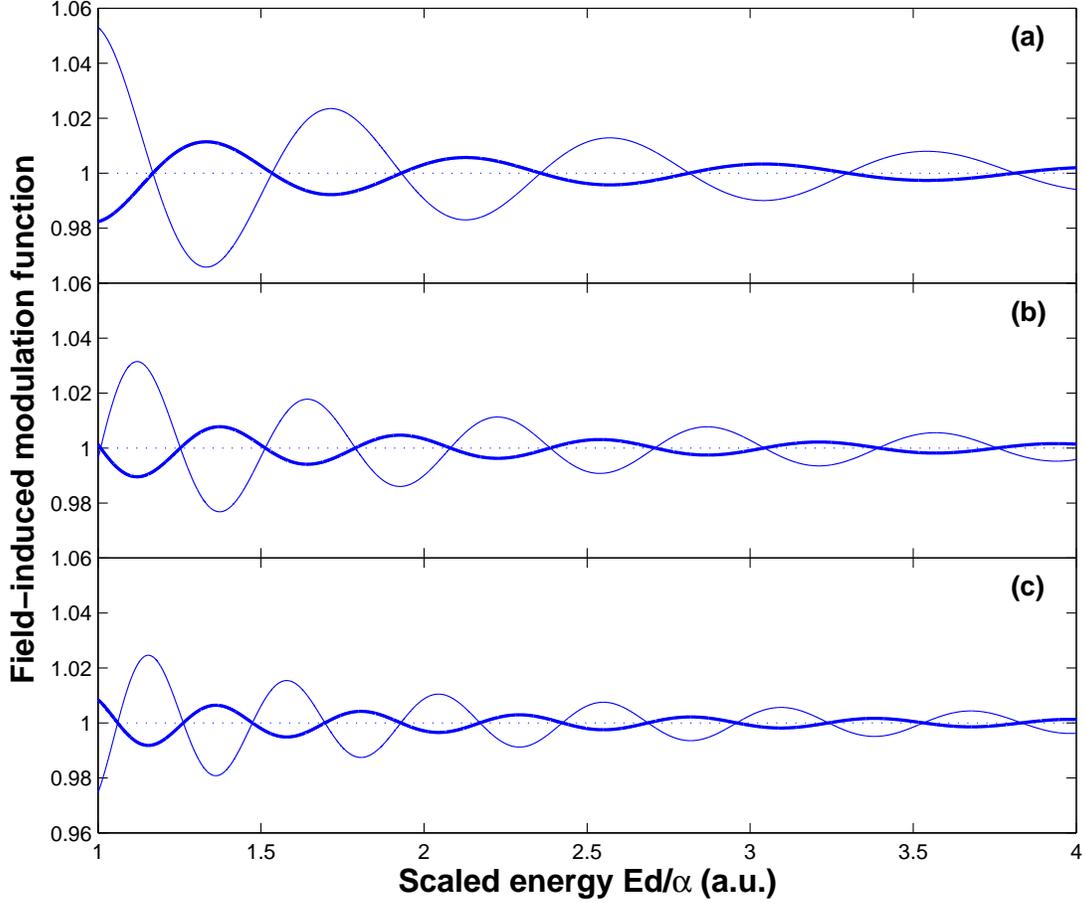}\\
  \caption{The field-induced modulation function $\mathcal{H}^c$  vs the scaled energy $\overset{\sim}E=Ed/\alpha$.
   $\alpha=1$. Thick lines represent $s$-wave photodetachment.
   Thin lines represent $p$-wave photodetament with linear polarization along $z$ axis.
  (a) $d=100\,a_0$, (b) $d=200\,a_0$, (c) $d=300\,a_0$.
  $a_0$ is the Bohr radius.
  A phase difference $\pi$ always exists.
  Increasing $d$ leads to the increase of oscillation frequency and decrease of amplitude.}
\end{figure}

\begin{figure}
  \centering
  \includegraphics[scale=1]{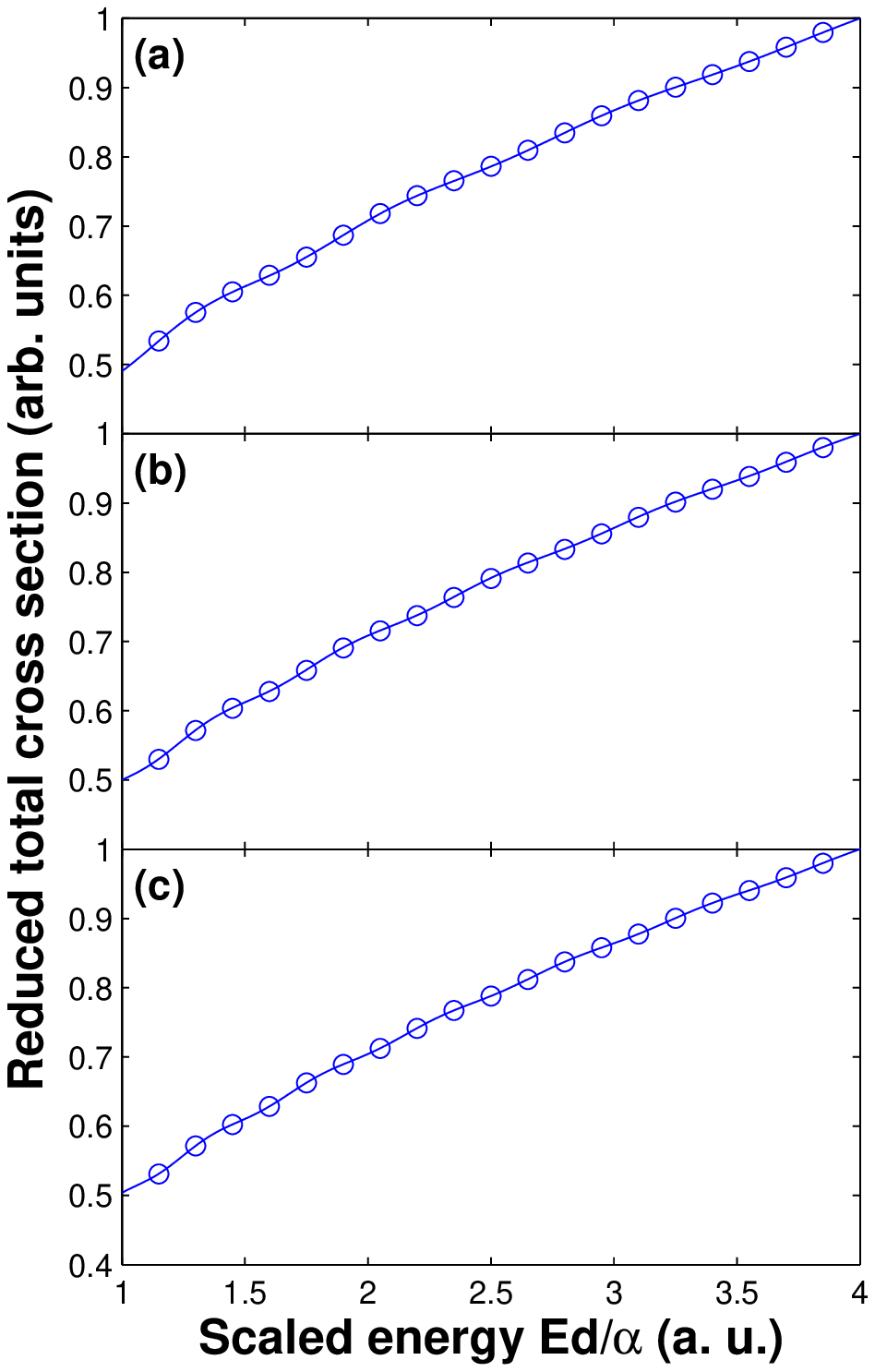}\\
  \caption{The reduced total cross section $\overset{\sim}{\sigma}$ vs the scaled energy $\overset{\sim}E=Ed/\alpha$
  for $s$-wave photodetachment. $\alpha=1$.
  (a) $d=100\,a_0$, (b) $d=200\,a_0$, (c) $d=300\,a_0$.
  $a_0$ is the Bohr radius.
  The open circles represent results obtained by integrating the differential cross section.}
\end{figure}

\begin{figure}
  \centering
  \includegraphics[scale=1]{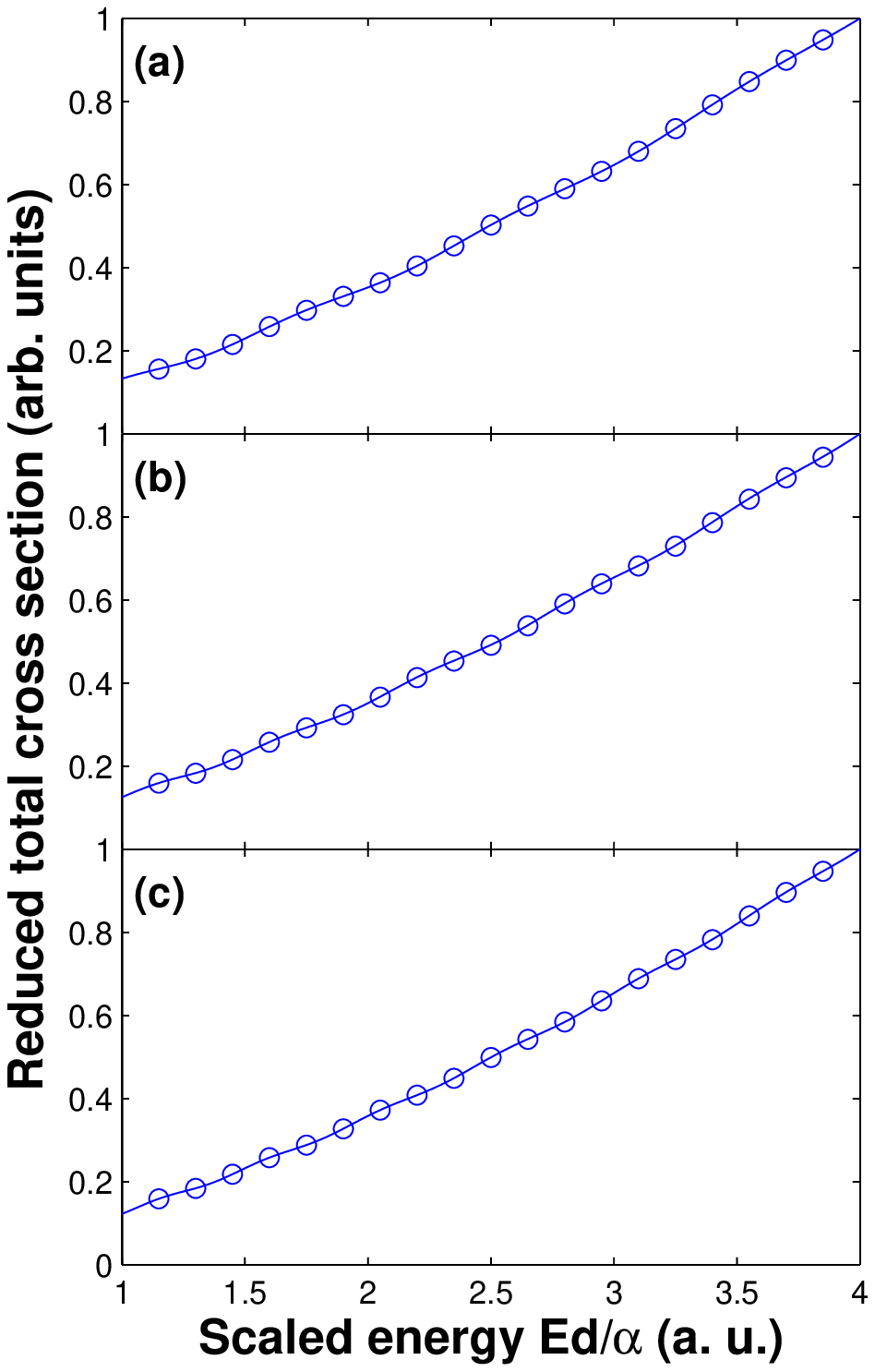}\\
  \caption{The reduced total cross section $\overset{\sim}{\sigma}$ vs the scaled energy $\overset{\sim}E=Ed/\alpha$
  for $p$-wave photodetachment with linear polarization along $z$ axis. $\alpha=1$.
  (a) $d=100\,a_0$, (b) $d=200\,a_0$, (c) $d=300\,a_0$.
  $a_0$ is the Bohr radius.
  The open circles represent results obtained by integrating the differential cross section.}
\end{figure}

\begin{figure}
  \centering
  \includegraphics[scale=.8]{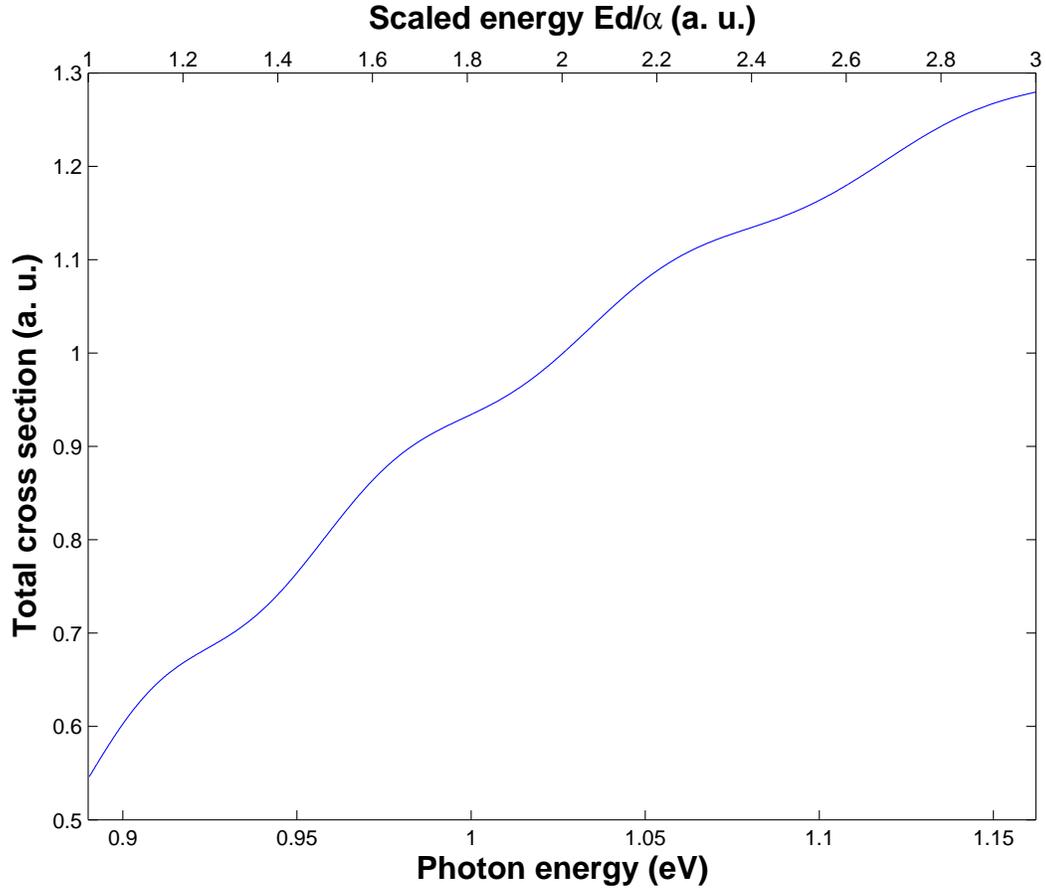}\\
  \caption{The total photodetachment cross section of $\rm{H^-}$ near an attractive center.
    The distance between the source point and the force center $d$ is $200\,a_0$,
  the positive charge number $\alpha$ is $1$.
  $a_0$ is the Bohr radius.
  The oscillatory structure is visible.}
\end{figure}

\begin{figure}
  \centering
  \includegraphics[scale=.8]{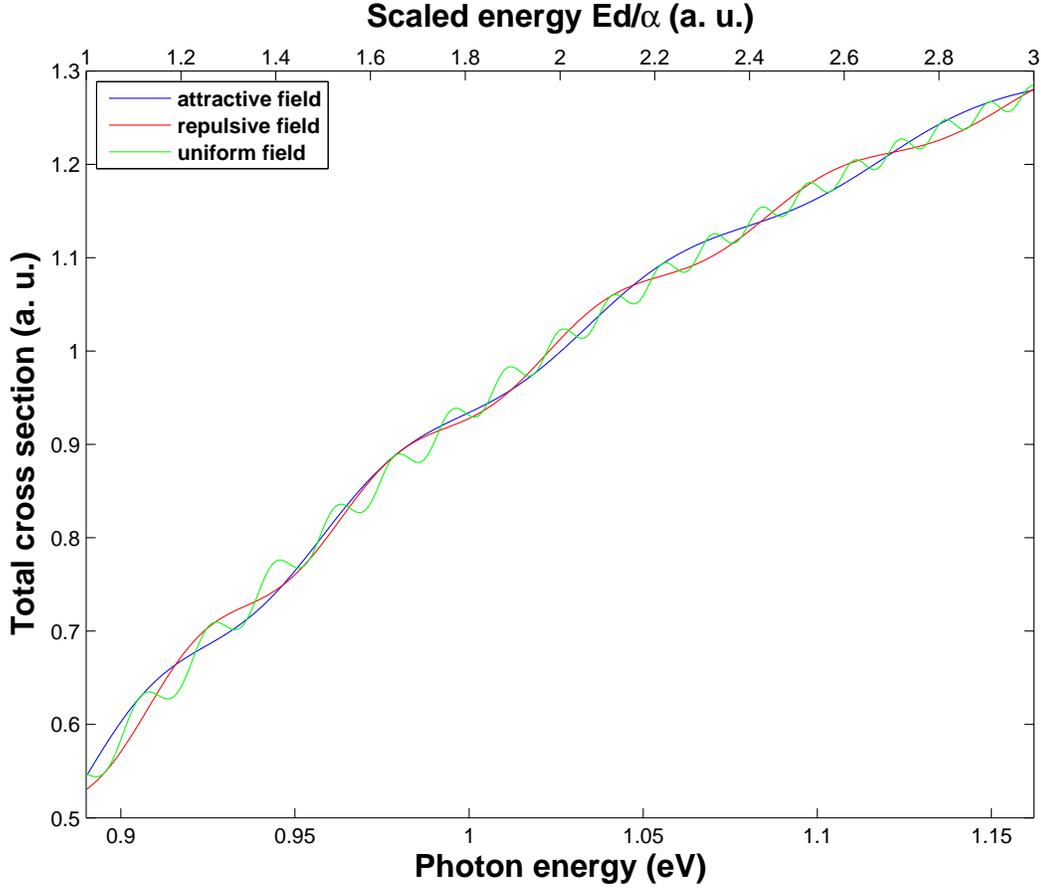}\\
  \caption{Comparisons of the total photodetachment cross section of $\rm{H^-}$
  near an attractive force center (blue),
  near a repulsive force center (red)
  and in a uniform electric field (green).
  The oscillation amplitudes are identical.
  The frequency is highest for the uniform field followed by the repulsive central field and the attractive field.
  The distance $d$ between the source center and the force (attractive or repulsive) center is $200\,a_0$.
  The charge number of the force center (attractive or repulsive) $\alpha$ is 1.
  The uniform field strength is $\alpha/d^2$ and equals $128.55\,\rm{kV/cm}$ for the above set.}
\end{figure}

\begin{figure}
  \centering
  \includegraphics[scale=0.8]{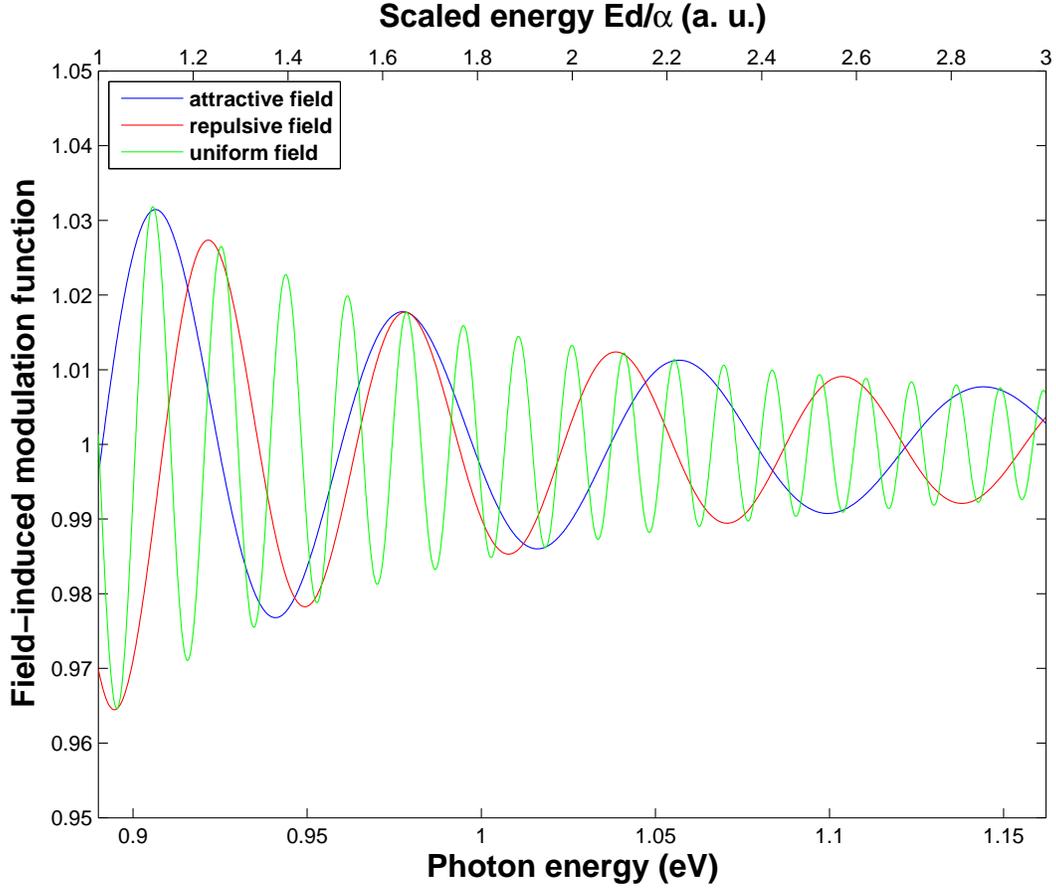}\\
  \caption{Comparisons of the modulation function for the photodetachment of an anion
  near an attractive force center (blue),
  near a repulsive force center (red)
  and in a uniform electric field (green).
  The oscillation amplitudes are the same.
  The frequency is highest for the uniform field followed by the repulsive central field and the attractive field.
  The distance $d$ between the source center and the force (attractive or repulsive) center is $200\,a_0$.
  The charge number of the force center (attractive or repulsive) $\alpha$ is 1.
  The uniform field strength is $\alpha/d^2$ and equals $128.55\,\rm{kV/cm}$ for the above set.}
\end{figure}

\begin{figure}
  \centering
  \includegraphics[scale=0.8]{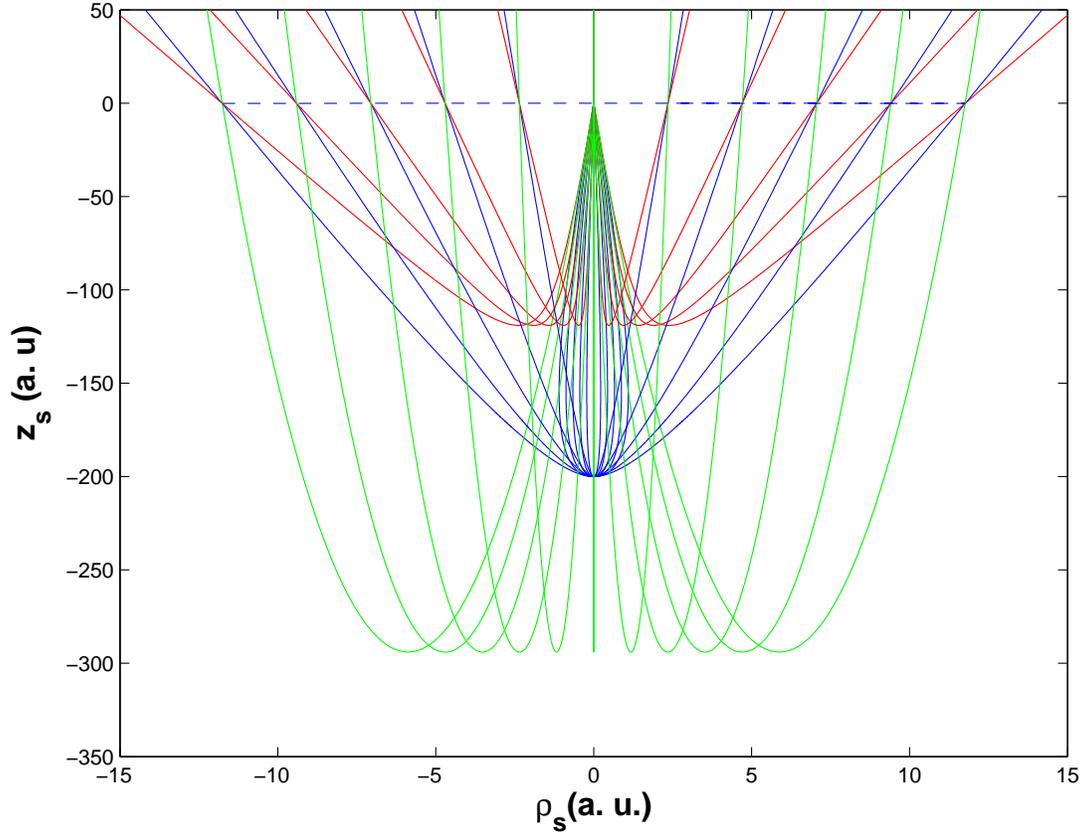}\\
  \caption{Electron trajectories near $\beta=\pi$ for the three systems:
   the photodetachment of an anion near an attractive force center (blue),
  near a repulsive force center (red)
  and in a uniform electric field (green).
  $d=200\,a_0$ , $\alpha=1$ and $E=0.2\,\rm{eV}$.
  The uniform field strength is $\alpha/d^2$ and equals $128.55\,\rm{kV/cm}$ for the above set.
  The trajectories with the same $\beta$ for the three systems cross each other when go back to the source region.
   Linking the intersections gives the dash line.}
\end{figure}

\clearpage

\begin{figure}
  \centering
  \includegraphics[scale=0.8]{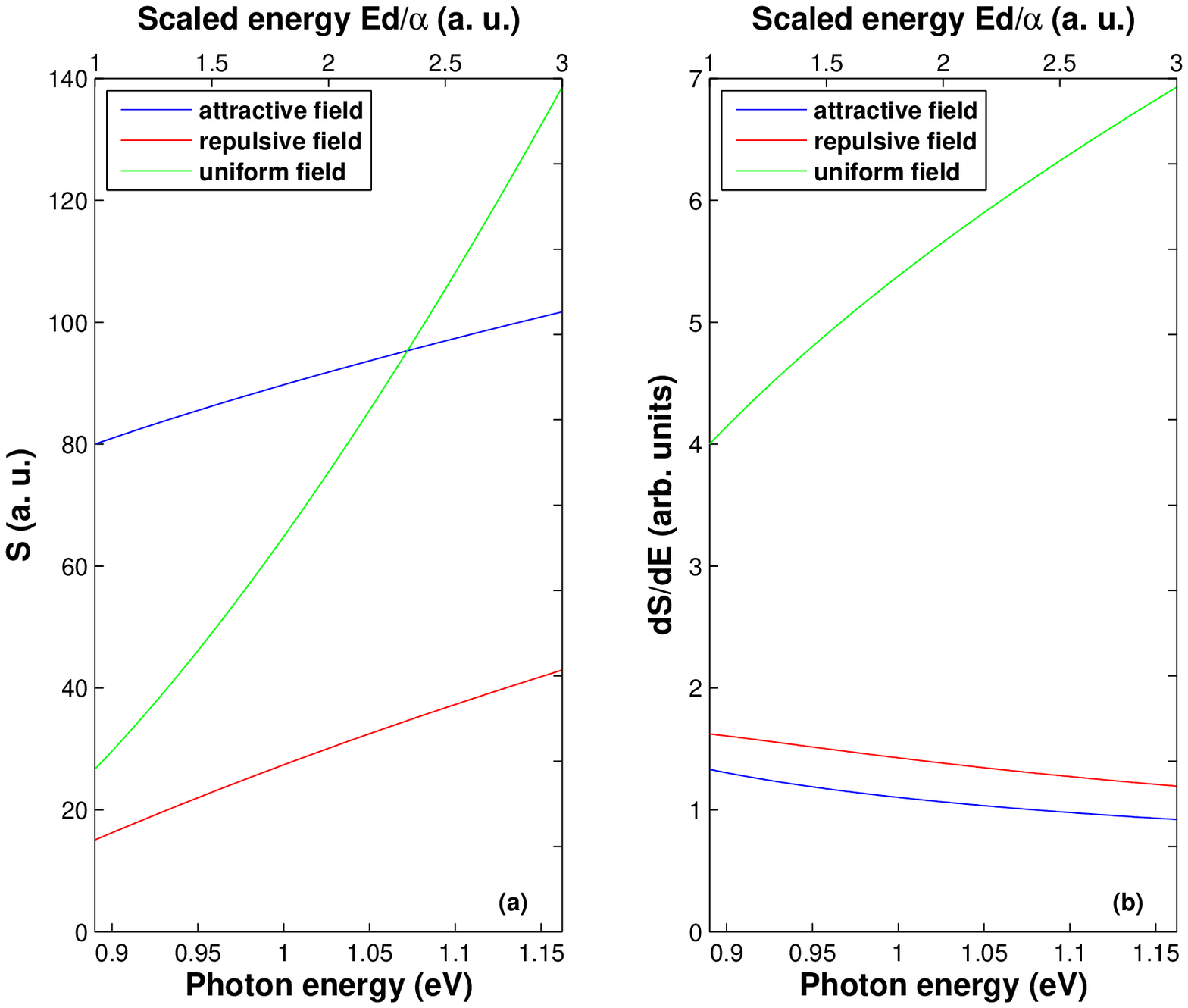}\\
  \caption{$d=200\,a_0$ and $\alpha=1$. (a) The phase $S$ vs the photon energy.
  (b)The differential of phase $S$ with respect to energy $E$ vs the photon energy.}
\end{figure}



\begin{references}

\bibitem{Demkov}
Yu. N. Demkov, V. D. Kondratovich, and V. N. Ostrovskii, JETP Lett. {\bf34}, 403 (1981)
\bibitem{Fabrikant1} I. I. Fabrikant, Sov. Phys.-JETP {\bf{52}}, 1045 (1980).
\bibitem{Bryant}
H. C. Bryant, A. Mohagheghi, J. E. Stewart, J. B. Donahue, C. R. Quick, R. A. Reeder,
V. Yuan, C. R. Hummer, W. W. Smith, Stanley Cohen, William P. Reinhardt, and Lillian Overman, Phys. Rev. Lett. {\bf58}, 2412 (1987).
\bibitem{Bryant2}
J. E. Stewart, H. C. Bryant, P. G. Harris, A. H. Mohagheghi, J. B. Donahue, C. R. Quick, R. A. Reeder,
 V. Yuan, C. R. Hummer, W. W. Smith, and Stanley Cohen, Phys. Rev. A {\bf38}, 5628 (1988)
\bibitem{Rau}
A. R. P. Rau and Hin-Yiu Wong, Phys. Rev. A {\bf 37}, 632 (1988).
\bibitem{Rau2}
 A. R. P. Rau and Chitra Rangan,
Phys. Rev. A {\bf 64}, 037402 (2001).
\bibitem{Du1} M. L. Du and J. B. Delos, Phys. Rev. A
{\bf 38}, 5609 (1988)
\bibitem{Du2}
M. L. Du, Phys. Rev. A {\bf 40}, 4983 (1989).
\bibitem{Du3}M. L. Du, Phys. Rev. A {\bf 70}, 055402 (2004).
\bibitem{Gibson}
 N. D. Gibson, B. J. Davies, and D. J. Larson, Phys. Rev. A {\bf 47},
1946 (1993).
\bibitem{Gibson2}
 N. D. Gibson, B. J. Davies, and D. J. Larson, Phys. Rev. A {\bf 48}, 310 (1993).
 \bibitem{Du4}
 M. L. Du, Phys. Rev. A {\bf 40}, 1330 (1989).
\bibitem{Peters1}
 Aaron D. Peters and John B. Delos, Phys. Rev. A {\bf
47}, 3020 (1993); {\bf 47}, 3036 (1993).
\bibitem{Peters2}
 Aaron D. Peters, Charles Jaff$\acute{\textrm{e}}$, and John B. Delos, Phys. Rev. A {\bf
56}, 331 (1997); Phys. Rev. Lett. {\bf 73}, 2825 (1994).
\bibitem{Xiaopeng Xing}
Xiao-Peng Xing, Xue-Bin Wang, and Lai-Sheng Wang, Phys. Rev. Lett. {\bf101}, 083003 (2008); J. Chem. Phys. {\bf130},
074301 (2009)
\bibitem{B. C. Yang}
B. C. Yang, J. B. Delos, and M. L. Du, Phys. Rev. A {\bf88}, 023409 (2013)
\bibitem{B. C. Yang2}
B. C. Yang, J. B. Delos, and M. L. Du, Phys. Rev. A {\bf89}, 013417 (2014)
\bibitem{Swenson}
J. K. Swenson, J. Burgd$\rm{\ddot{o}}$fer, F. W. Meyer, C. C. Havener, D. C. Gregory, and N. Stolterfoht,
Phys. Rev. Lett. {\bf66}, 417 (1991)
\bibitem{Du and Delos}
M. L. Du and J. B. Delos, Phys. Rev. Lett. {\bf58}, 1731 (1987);
Phys. Rev. A {\bf38}, 1896 (1988); {\bf38}, 1913 (1988);
\bibitem{Gutzwiller}
Martin. C. Gutzwiller, J. Math. Phys. {\bf12}, 343 (1971).
\bibitem{Delos}
J. B. Delos, Adv. Chem. Phys. {\bf65}, 161 (1986).

\bibitem{Blondel1}
Chrisophe Blondel, Christian Delsart and Francois Dulieu, Phys. Rev. Lett. {\bf77}, 3755
(1996).
\bibitem{Blondel2}
Micka$\ddot{e}$l Vandevraye, Philippe Babilotte, Cyril Drag, and Christophe Blondel, Phys. Rev. A {\bf90}, 013411
(2014).
\bibitem{Du6}
M. L. Du, Eur. Phys. J. D {\bf38}, 533 (2006)






\end{references}
\end{document}